\documentclass[prd,amsmath,showkeys]{revtex4}
\pdfoutput=1 
\usepackage[usenames]{color}
\usepackage{graphicx}
\usepackage{bm}

\def\sst{\scriptscriptstyle}

\def\LQCD{{{\cal L}_{\sst{QCD}}}}

\def\paru{{\partial_\mu}}
\def\parv{{\partial_\nu}}
\def\Dux{{\Delta u(x)}}
\def\Ddx{{\Delta d(x)}}
\def\DS{{\Delta \Sigma}}
\def\DG{{\Delta G}}
\def\Fmn{{F^{\mu\nu}}}
\def\Gaup{{G^a_{\mu\nu}}}
\def\Gadn{{G_a^{\mu\nu}}}
\def\Gamu{{G^a_\mu}}
\def\Ganu{{G^a_\nu}}
\def\Gbmu{{G^b_\mu}}
\def\Gcnu{{G^c_\nu}}
\def\fabc{{f_{\sst{abc}}}}
\def\be{\begin{equation}}
\def\ee{\end{equation}}
\def\bea{\begin{eqnarray}}
\def\eea{\end{eqnarray}}
\def\qv{{q_{\scriptscriptstyle V}}}
\def\uv{{u_{\scriptscriptstyle V}}}
\def\dv{{d_{\scriptscriptstyle V}}}
\def\pT{{p_{\scriptscriptstyle T}}}
\def\RAx{{R_{\scriptstyle A}(x)}}
\def\als{{\alpha_s}}
\def\IE{{\it{i.e.}}}
\def\EG{{\it{e.g.}}}
\def\EA{{\it{et al.}}}

\begin{document}

\title{Nucleon Resonances and Quark Structure}

\author{J.T. Londergan\\
Physics Dept and Nuclear Theory Center, Indiana 
University, \\ Bloomington, IN 47405, USA.\\
tlonderg@indiana.edu}


\date{\today}

\begin{abstract}
A pedagogical review of the past 50 years of study of resonances, leading to 
our understanding of the quark content of baryons and mesons. The level of 
this review is intended for undergraduates or first-year graduate students.  
Topics covered include: the quark structure 
of the proton as revealed through deep inelastic scattering; structure 
functions and what they reveal about proton structure; and prospects for 
further studies with new and upgraded facilities, particularly 
a proposed electron-ion collider.
\end{abstract}

\maketitle

\section{Introduction}
\label{Sec:Intro}

These lectures were given at Aligarh Muslim University, Aligarh India, at a 
nuclear physics workshop attended by Indian graduate students and 
postdocs. These lectures provide a pedagogical review of the past 
50 years of experimental and theoretical studies of the 
quark and gluon structure of the nucleon. Since many speakers at this 
workshop will discuss structure functions extracted from deep inelastic 
scattering reactions, we define these quantities and review what can be 
extracted from them. We hope that specialists in this field will tolerate 
our brief and superficial summary of many years' study of strongly-interacting 
systems. 

\begin{figure}[ht]
\includegraphics[width=1.2in,angle=0]{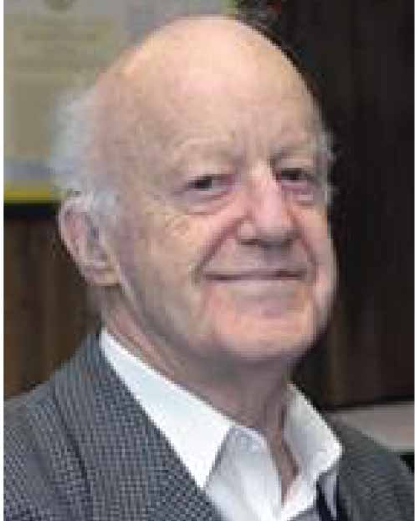}
\caption{\label{Fig:Dalitz} Richard H. Dalitz, 1925-2006.}
\end{figure}

These lectures are dedicated to my mentor, Richard Dalitz, who was from 
1963-2000 Royal Society Professor of Physics at Oxford 
University and who made seminal contributions in particle 
phenomenology. Dalitz pairs (electron-positron 
pairs from decay of a high-energy photon) and the Dalitz plot were named 
after him. In addition he made important contributions to our 
understanding of strange particles and hypernuclei, and to  
particle spectroscopy through the constituent-quark model. I appreciated his 
insights into theoretical physics, his keen intuition about  
physical problems, and his encouragement of young physicists.  

\section{Discovery of New Particles with Dedicated Accelerators}

The development of both accelerators and detectors progressed rapidly 
through the first half of the $20^{th}$ century \cite{Clo02}. Ernest 
Lawrence's cyclotron ushered in a new era which provided both much higher 
beam energies and more intense beams than were previously possible. This was 
followed by the development of synchrotrons and mastery of 
magnetic focusing techniques, which allowed particles to be 
contained in a circular pipe surrounded by individual focusing magnets. 
Shortly following World War II, development began on a succession of machines 
based upon the synchrotron principle. Advances in accelerator technology were 
accompanied by corresponding advances 
in particle detectors. The bubble chamber was an extremely important tool 
in discovering new particles. In particular strange particles like the  
$\Xi$ and $\Omega$ were discovered in bubble chambers. Another major 
detector advance was the development of the spark chamber. 

Most of the particles discovered before the 1950s were either 
completely stable (so far as we know), like the electron or proton, or had 
relatively 
long lifetimes. For example, the lifetimes of charged pions and 
kaons are roughly $10^{-8}$ seconds, and the lifetimes of the $\Lambda$ and 
charged $\Sigma$ are of order $10^{-10}$ seconds. When charged (neutral) 
particles move with sufficiently large velocities, they leave measurable 
tracks (gaps) in emulsions or bubble chambers between their production and 
decay vertices. 

\begin{figure}[ht]
\includegraphics[width=4.4in,angle=0]{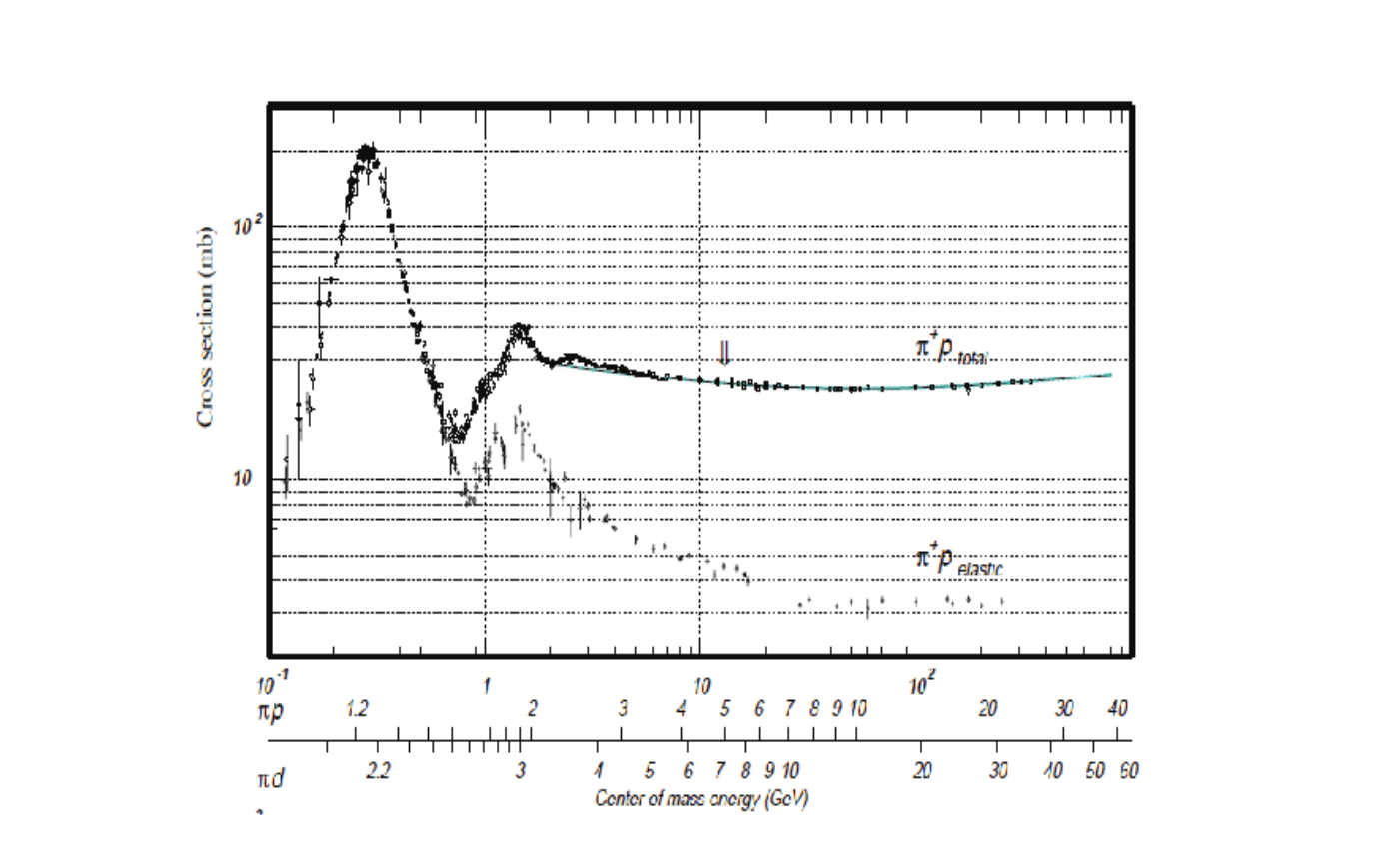}
\caption{\label{Fig:piPxsect} Total and elastic X-sections for $\pi^+-p$ 
scattering, as a function of the laboratory beam momentum in GeV/c (note 
logarithmic scale), also the CM energy in GeV.}
\end{figure}

Starting in the early 1950s, researchers began to discover 
states which were much more short-lived than these earlier particles. 
These states were extracted from scattering experiments at what were 
then the highest available energies. 
Fig.~\ref{Fig:piPxsect} shows elastic and total cross sections for $\pi^+-p$ 
scattering. Note the very prominent peak at CM energy just near 1.2 GeV. 
This is what was then called the `$P_{33}$ resonance' in $\pi-p$ 
scattering. The notation for this state denoted the angular momentum (the 
spectroscopic notation $P$ for $L = 1$), and the isospin $I$ and total spin 
$J$ in the notation $L_{2I,2J}$. In current terminology this is the 
$\Delta (1232)$ resonance \cite{PDG}. Since the total width for this state 
is of order 100 MeV then from the Heisenberg uncertainty principle which can 
be written as $\Delta E \Delta t \ge \hbar$, we can infer that the 
lifetime of the $\Delta (1232)$ is of order $10^{-23}$ seconds. 

For such short-lived states the production and decay vertices could never 
be separated, regardless of the particle's velocity. 
As scattering energies increased, and as nuclear targets were bombarded 
with different probes, more and more peaks appeared in cross sections. 
Determining the properties of these \textit{resonances} became a major area of 
research beginning about 1951, and spectroscopic studies of resonance 
production have continued ever since. Information about these resonances, 
and comparative ratings of the reliability of claimed resonant states, 
are compiled in the biennial report of the Particle Data Group \cite{PDG}. 

\subsection{Resonances and Scattering Amplitudes}
\label{Sec:resonances}

Here is an elementary review of bound states and scattering resonances 
\cite{Gol64}. Consider a two-body scattering process; for simplicity 
we will initially assume spinless non-relativistic particles in an 
$S$-wave state, interacting via a potential $V(r)$. For the 
time being we will consider two-body scattering in only the elastic 
channel.  The radial wave function can be written as $R(r) = \psi(r)/r$, and  
the radial Schr\"odinger equation satisfied by $\psi(r)$ for a given value 
of the relative momentum $k$ is 
\be 
-\frac{d^2r \psi_k(r)}{dr^2} + U(r)\psi_k(r) = k^2\psi_k(r)
\label{eq:Schrod}
\ee
In Eq.~(\ref{eq:Schrod}), $U(r) = 2mV(r)$, $m$ is the reduced mass of the 
two-body pair, and we adopt units where $\hbar = 1$. 

Strong interaction potentials are short-ranged, so we expect $U(r) \rightarrow 
0$ as $r \rightarrow \infty$.  In this case at large $r$ we can write 
\be 
\lim_{r \rightarrow \infty} \ \ \psi_k(r) = a(k) e^{-ikr} + b(k) e^{ikr}
\label{eq:psilim}
\ee
We can consider the wavefunction $\psi$ to be an analytic function of $k$.  
A sufficiently strong attractive potential may support one or more 
bound states. These are confined states that decay exponentially. They will 
occur as discrete solutions of the Schr\"odinger equation of 
Eq.~(\ref{eq:Schrod}) corresponding to imaginary momentum $k = ik_B$.  
For this particular value of the momentum, the coefficient $a(ik_B)$ in 
Eq.~(\ref{eq:psilim}) must vanish; in this case the 
wavefunction will decay exponentially at large distances, 
\IE~$\psi_{k_B}(r) \sim  e^{-k_Br}$. 
 
The properties of bound states and resonances are related to  
the behavior of the scattering amplitude in the complex $k$ plane, or 
equivalently the complex $E$ plane. The non-relativistic relation between 
energy and momentum is 
\be
E = k^2/(2m) + E_{th}
\label{eq:Evsk}
\ee
From Eq.~(\ref{eq:Evsk}) it is clear that as the phase of the complex 
variable $k$ varies from $0 \rightarrow 2\pi$, the phase of $E$ will 
change from $0 \rightarrow 4\pi$. Thus one sheet of 
the complex $k$ plane will map onto two sheets of the complex $E$ plane. For 
the time being consider only the elastic channel (inelastic channels will 
produce additional sheets of the complex $E$ plane). 

\begin{figure}[ht]
\includegraphics[width=6.0in,angle=0]{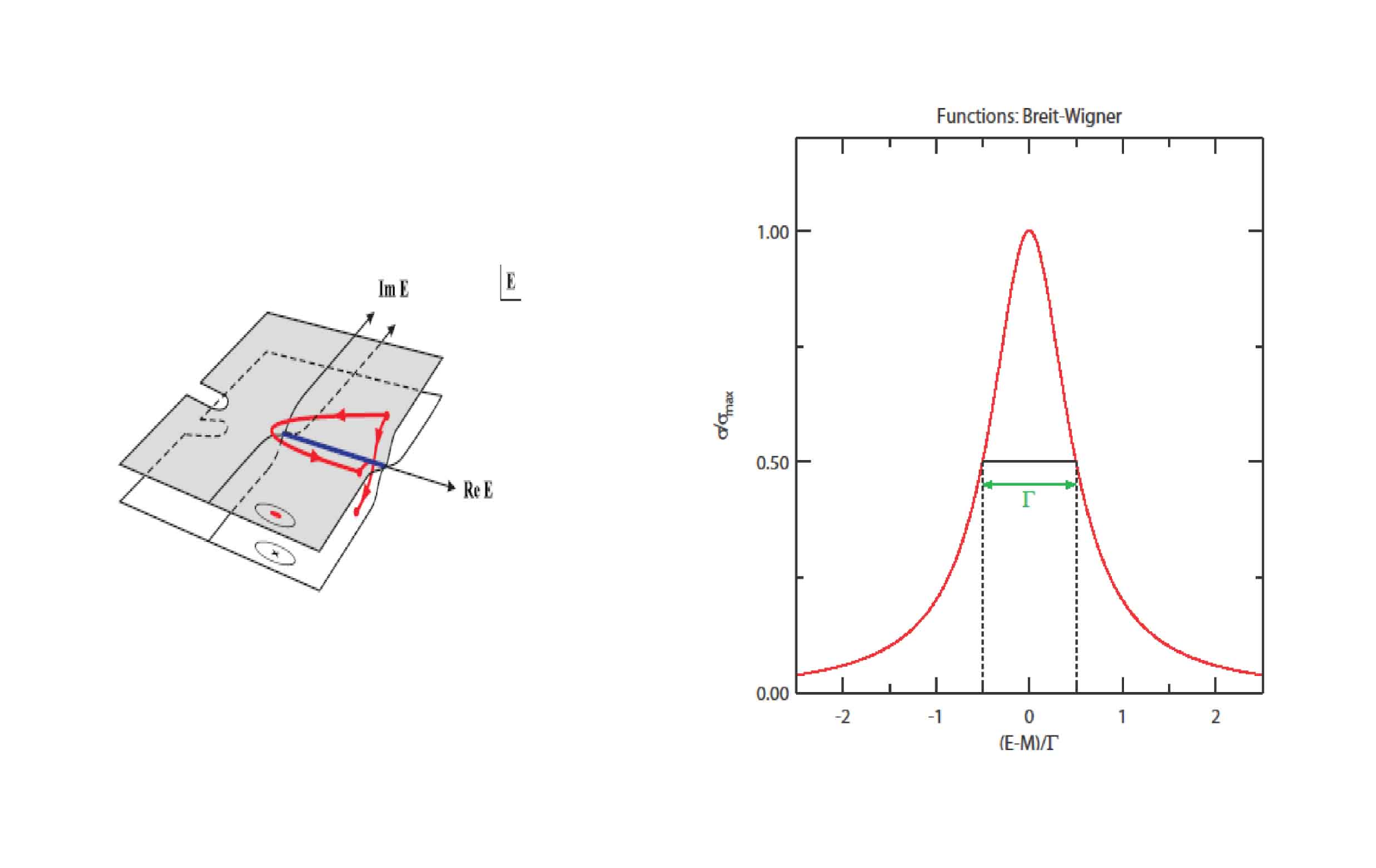}
\vspace{-1.0cm}
\caption{\label{Fig:Ecmplx} Left: the complex $E$ plane, the physical sheet 
(top,shaded) and an unphysical sheet (bottom). A cut runs along the real $E$ 
axis, beginning at continuum threshold. A resonance is shown as an 'X' on the 
unphysical sheet. Right: X-section vs.\ energy for 
a Breit-Wigner resonance, whose amplitude is given by 
Eq.~(\protect{\ref{eq:BW}}).}
\end{figure}

A bound state corresponds to a pole of the scattering amplitude 
${\mathcal T}(E)$ in the complex $k$ plane at positive imaginary value 
$ik_B$. From Eq.~(\ref{eq:Evsk}), bound state poles will appear 
on the real $E$ axis, below the continuum threshold $E_{th}$.  
The upper sheet of the complex $E$ plane is called the physical sheet (the 
shaded sheet on the left in Fig.~\ref{Fig:Ecmplx}), and the lower 
sheet is an unphysical sheet. The two sheets are joined by a cut running 
along the positive real $E$ axis, beginning at continuum threshold.   
 
If one crosses the cut starting from above, moving from the physical sheet 
onto the unphysical sheet, the ${\mathcal T}$ matrix is continuous. However,  
${\mathcal T}$ is discontinuous if one crosses the cut beginning and remaining 
on the physical sheet. The discontinuity of the ${\mathcal T}$ matrix 
across the cut plays an important role in constructing dispersion relations 
for the scattering amplitude \cite{Gol64}.  

Imagine that one has an attractive potential strong enough to 
produce a bound state. As one decreases the strength of the attractive 
potential, at some point it will no longer 
support a bound state. The bound state pole may move onto 
the second sheet of the complex $E$ plane and become a resonance. 
A resonance generally produces a peak in 
the scattering cross section at some energy above elastic threshold. 
The prototypical non-relativistic resonance amplitude was posited by Breit 
and Wigner \cite{Bre37}, who wrote the scattering amplitude in the vicinity 
of the resonance in the form  
\be
 {\cal T}(E) = g(E)[E- M + i\Gamma/2]^{-1}
\label{eq:BW}
\ee
This leads to a peak in the cross section vs.\ energy for $E \sim M$, 
a so-called `Breit-Wigner' resonance. The right-hand side of 
Fig.~\ref{Fig:Ecmplx} plots the 
qualitative features of the resulting cross section, which is proportional 
to the absolute square of the scattering amplitude. 

Eq.~(\ref{eq:BW}) shows that the resulting cross section 
will have a peak at energy $E=M$; furthermore, if the quantity $g(E)$ is 
relatively constant in the vicinity of the peak, then the full width at 
half maximum around the peak will be given by the quantity $\Gamma$. So 
the location and width of the resonance peak are characterized by the 
two quantities $M$ and $\Gamma$. Eq.~(\ref{eq:BW}) shows that the scattering 
amplitude will have a pole in the complex plane at $E = M - i\Gamma/2$.  
The responance pole lies on the 
unphysical sheet of the complex $E$ plane, below the real $E$ axis, 
as indicated on the left in Fig.~\ref{Fig:Ecmplx}.   

We can also examine the properties of resonances by defining the 
scattering or ${\mathcal S}$-matrix \cite{Gol64}. The ${\mathcal T}$ matrix 
is defined as 
the amplitude of an outgoing wave solution, corresponding to a unit incoming 
wave. If we restrict ourself to the elastic channel, the  
relation between the ${\mathcal S}$ and ${\mathcal T}$ matrix is 
\bea
{\cal T}(E) &=& \frac{S(E) - 1}{2i}; \hspace{0.4cm}
S(E) = e^{2i\delta(E)}; \hspace{0.4cm} 
{\cal T}(E) = e^{i\delta(E)}\sin(\delta(E))
\label{eq:Smtrx}
\eea
Unitarity requires that the $S$-matrix have absolute value one. An 
attractive (repulsive) potential tends to produce positive (negative) 
phase shifts. In a one-channel situation with a Breit-Wigner 
resonance, Eq.~(\ref{eq:BW}) shows that 
${\mathcal T}(E)$ will be purely imaginary at the resonance position 
$E=M$. By inspection of Eq.~(\ref{eq:Smtrx}), a purely imaginary 
${\mathcal T}$-matrix corresponds to $\delta(E=M) = \pi/2$. 
The location of an elastic resonance (with no 
background) is characterized by a phase shift of $90^\circ$.  

\begin{figure}[ht]
\vspace{-1.0cm}
\includegraphics[width=3.0in,angle=0]{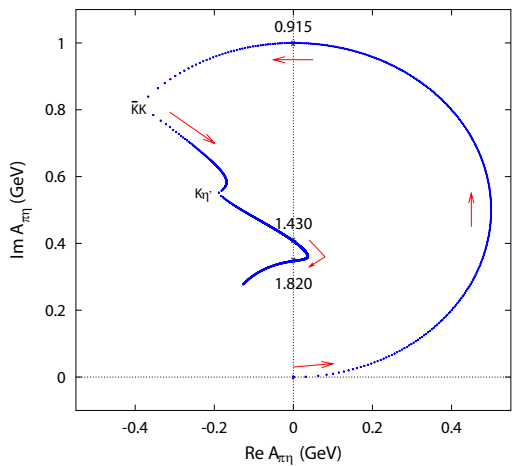}
\vspace{-1.5cm}
\caption{\label{Fig:Argand} Argand diagram. At low energies the 
${\mathcal T}$ matrix moves on the unitary circle. The phase $\delta = \pi/2$ 
at an elastic resonance. When inelastic channels open, the ${\mathcal T}$ 
matrix moves inside the unitary circle.}
\end{figure}

An Argand diagram represents a convenient graphical way to study the 
behavior of the scattering amplitude, through a  
two-dimensional plot of the ${\mathcal T}$ matrix vs.~energy. The 
horizontal axis is the real part of the ${\mathcal T}$ matrix and the 
vertical axis is the imaginary part. For  
purely elastic scattering the ${\mathcal T}$ matrix moves on a circle 
(the unitary circle) of radius $1/2$ centered at the point $(0,1/2)$. 
The origin of the Argand diagram corresponds to $\delta = 0$. 
The angle subtended by any point on the unitary circle (relative to the 
origin) is $2\delta$. This is illustrated in  
Fig.~\ref{Fig:Argand}, which shows an Argand diagram for $\pi-\eta$ 
scattering \cite{Bog02}. For sufficiently low 
energies, only the elastic channel is open, and the ${\mathcal T}$  
matrix moves on the unitary circle. The phase shift $\delta = \pi/2$ at the 
position of the $a_0$(980) resonance. The smaller the width $\Gamma$, the 
faster the phase shift will move through the resonance.  

At higher energies, inelastic channels open up. The 
complex $E$ plane must be expanded to include inelastic channels. 
Every time an inelastic channel opens, a new cut appears on the real $E$ 
axis at the inelastic channel threshold, and new unphysical sheets appear. 
Locating a scattering resonance becomes more difficult, as one 
must determine both the resonance position and the sheet on which 
it occurs. In the presence of inelastic channels the scattering amplitude 
becomes a matrix.  The quantity 
${\mathcal T}_{fi}(E)$ is defined as the coefficient of an outgoing wave 
in channel $f$ corresponding to unit incoming wave in channel $i$. Unitarity 
requires that in the presence of inelastic channels, the elastic 
${\mathcal T}$ matrix element must lie inside the unitary circle, and 
Fig.~\ref{Fig:Argand} demonstrates this. Finding 
resonances in multichannel reactions is likely to rely on 
sophisticated fitting of scattering amplitudes, or on models 
that allow one to analytically continue the 
${\mathcal T}$ matrix in the complex $E$ plane.     

We should emphasize that the simple Breit-Wigner resonance is almost never 
observed in actual experiments. The $\pi-\pi$ resonance $\rho$(770) is one of 
the few examples where this simple resonance picture is valid \cite{PDG}. In 
most other cases the situation is far more complex. The goal in analyzing 
resonant reactions is to identify the location of the pole in the complex 
energy plane, as this pole is the fundamental dynamical quantity. 

\subsection{Identification of Resonances with the Dalitz Plot}
\label{Sec:Dalitz}

In the 1950s the catalog of resonances 
grew at first modestly, then exploded. From a few resonances, one rapidly 
accumulated a dozen, then eventually greater than one hundred resonant 
states! A major contributor to this `zoo' of resonant states lay in the 
ability for accelerators to produce secondary beams of pions and kaons. 
These could be scattered from protons and nuclei, and the final states 
could be analyzed. 

Our simple analysis of resonances was carried out for spinless 
particles. For particles with spin it is necessary to combine spin and 
orbital angular 
momentum in order to determine the total spin and parity of a given 
resonance. Sophisticated multi-channel partial-wave analyses were developed to 
extract the scattering amplitudes from experimental reaction data. As energies 
increased many final states involved three or more strongly interacting 
particles. A major advance in resonance phenomenology was 
the introduction of the \textit{Dalitz plot}, a 
method to analyze three-body (and higher) final states from 
scattering reactions \cite{Dal54}. 

\begin{figure}[ht]
\vspace{-1.0cm}
\includegraphics[width=3.3in,angle=0]{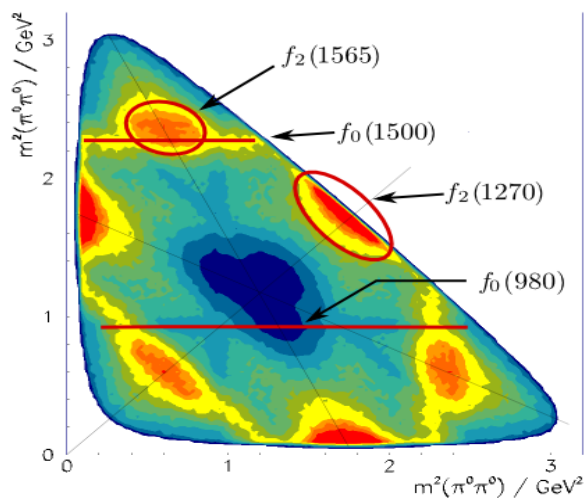}
\vspace{-1.0cm}
\caption{\label{Fig:Dalitzplot} Dalitz plot for the reaction $p + \bar{p} 
\rightarrow 3\pi^0$. The shading 
represents the frequency of events, with blue denoting few and red 
signifying many events. Four resonances are visible in this plot.}
\end{figure}

To understand the Dalitz plot, consider a reaction $a + b \rightarrow 1 + 
2 + 3$ leading to three final-state particles. For the Dalitz plot one graphs 
the data vs.~the squared invariant mass of one pair of final-state particles 
on one axis and the invariant mass of a second pair on a perpendicular axis. 
The data points fall inside well defined kinematic limits, since one can show  
$m_{12}^2 + m_{13}^2 + m_{23}^2 =$ const. Each observed event corresponds to 
a point inside the Dalitz plot. If the scattering state decays 
uniformly into three particles, the resulting points will be uniformly 
distributed. However, two-body resonances appear as enhancements in the 
Dalitz plot, at the location of the resonant mass. Furthermore, the structure 
of these bands depends on the spin and parity of the resonant state(s). 

Fig.~\ref{Fig:Dalitzplot} shows the Dalitz plot for the reaction $p + \bar{p} 
\rightarrow 3\pi^0$, from the Crystal Barrel experiment \cite{Ams02}. The 
data points are 
represented by shading, with blue denoting few and red 
signifying many events. The axes are chosen as the invariant mass of two 
different $\pi^0$ pairs. The pion Bose symmetry requires that the Dalitz plot 
be symmetric about any axis that exchanges two pions. Four resonances 
are identified in this plot. The spin-0 $f_0$(1500) appears as 
a simple enhancement. Two spin-2 resonances, the $f_2$(1270) 
and $f_2$(1565), have a more complicated shape. The final resonance, the 
$f_0$(980) with spin $0$, also appears on the plot. Because of quantum 
mechanical interference between amplitudes, the 
$f_0$(980) appears as a \textit{dip} rather than a peak in the distribution 
of events. The Dalitz plot has proved incredibly useful in enabling 
experimentalists to locate resonances in reactions with 
multi-particle final states. In addition, this technique provides   
a quick estimate of the spin and parity of the observed state.  

\section{The Quark Structure of Hadrons}
\label{Sec:Classify}

By the beginning of the 1960s, there were of order 100 known resonances. These 
could be classified into \textit{baryon resonances}, strongly interacting 
states with half-integral total angular momentum $J$, and 
\textit{meson resonances}, states with integral (or zero) spin. What had 
initially been exciting discoveries of a few new states had become almost an 
embarrassment. This apparent over-abundance of resonances was for awhile quite 
perplexing. The situation seemed to cry out for introduction of a 
classification scheme that would make sense of the masses and spins of 
the plethora of observed resonances. 

\begin{figure}[ht]
\includegraphics[width=5.0in,angle=0]{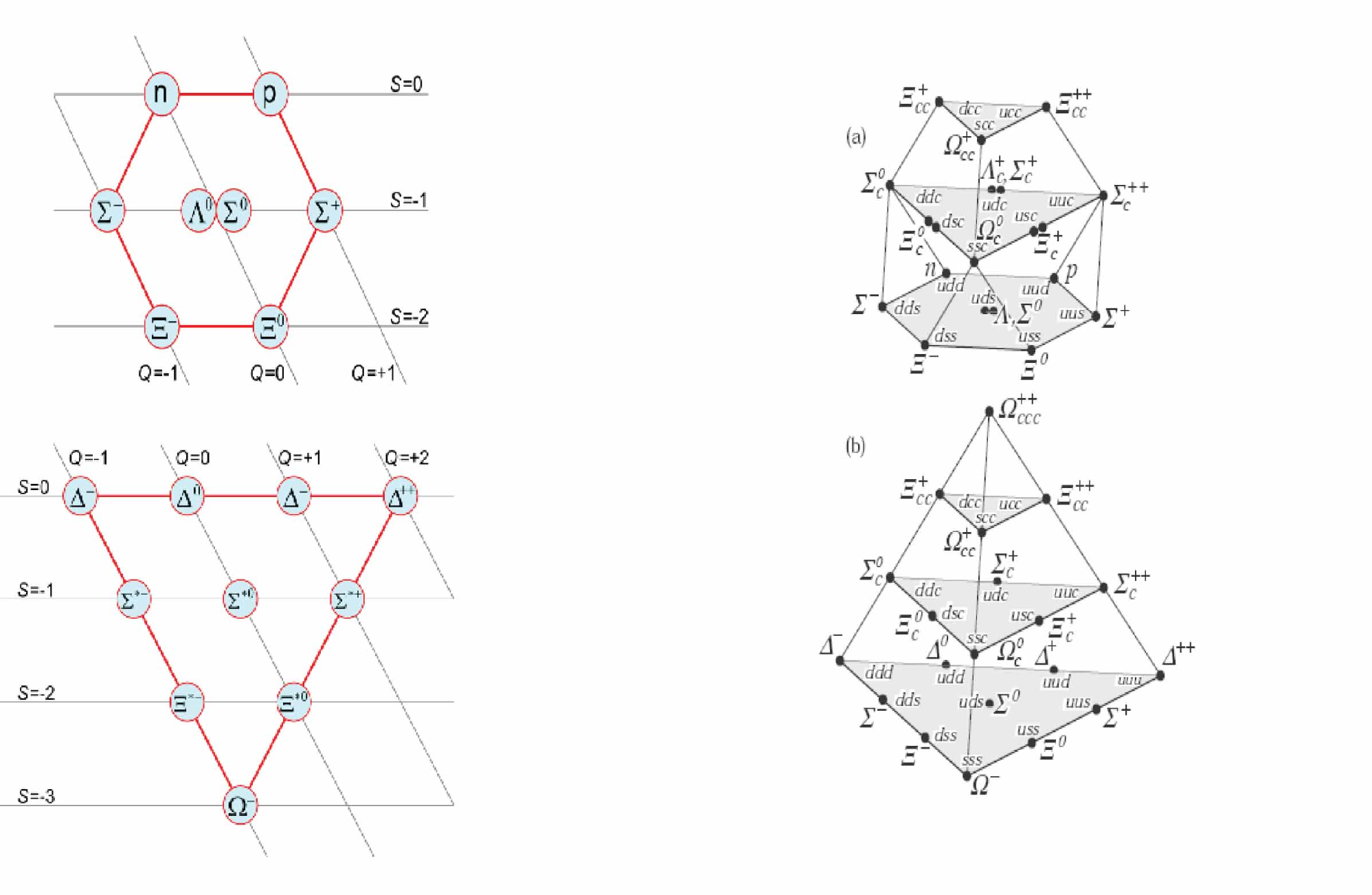}
\caption{\label{Fig:multiplets} Left: the baryon octet and decuplet as 
described by $SU(3)$ symmetry. The vertical axis is strangeness, the 
horizontal axis is the third component of isotopic spin, and lines of 
constant electric charge are indicated. Right: extensions of the octet 
and decuplet including charm quarks.}
\end{figure}

The first significant breakthrough came in 1960-61, when 
Ne'eman and Gell-Mann independently showed that nearly all mesons 
and baryons could be grouped in multiplets defined by the 
symmetry $SU(3)$ \cite{Gel64}. In this scheme, dubbed the `Eightfold Way' 
by Gell-Mann, 
the observed resonances occurred in families when plotted vs.~their 
electric charge, isotopic spin and strangeness values. 
The graph at left in Fig.~\ref{Fig:multiplets} shows the $SU(3)$ 
classification picture for the baryon octet (a multiplet of spin-parity 
$J^P= 1/2^+$ particles) and decuplet with $J^P =3/2^+$. 
The right-hand figure shows the extension of octet and decuplet 
multiplets when one includes charm quarks \cite{PDG}. 

The baryon octet consists of a pair of particles (the neutron and proton) with 
zero strangeness, four particles with $S=-1$ consisting of an isotropic 
triplet (the $\Sigma$s) and an isosinglet $\Lambda$, and a doublet of $\Xi$ 
with strangeness $S=-2$. One could derive 
formulae relating the masses of particles within a multiplet; these formulae 
were generally accurate to within a few percent. Significant support for this 
classification scheme came in 1962 with the discovery of the $\Xi^*$ doublet 
with strangeness $-2$ \cite{Sch63}. This left only one 'missing' 
member of the baryon decuplet, and Gell-Mann predicted a new baryon with 
spin-parity $3/2^+$ and strangeness $-3$ \cite{Gel64} dubbed the $\Omega^-$; 
multiplet mass formulae predicted a mass 1680 MeV. In 1964 the $\Omega^-$ 
was discovered at Brookhaven in bubble chamber photographs of $K^- + p$ 
reactions \cite{Bar64}; its mass was estimated as 1686 MeV. 

Similar $SU(3)$ multiplets, with corresponding mass formulae for the multiplet 
members, could be constructed for the known pseudoscalar and vector mesons. 
The next step was to find an underlying theory that explained why mesons 
and baryons should inhabit such multiplets. This answer was 
provided in 1964 independently by Gell-Mann \cite{Gel64b} and Zweig 
\cite{Zwe64}, who proposed that  
baryon and meson structure could be explained naturally by the existence of 
\textit{quarks} (Zweig called them \textit{aces}). The idea was that there 
existed three types of quarks, now called \textit{flavors} (up, down and 
strange), each with its corresponding antiquark. Baryons were composed of 
three quarks and a meson was a quark-antiquark pair. Individual quarks were 
fermions with intrinsic spin $1/2$; strangeness was an additive property, with 
the up and down quarks having $S=0$ and the strange quark $S=-1$. However, in 
order to produce the observed spectrum of particles, it was necessary that 
quarks possess electric charges that were a fraction of the proton's charge, 
specifically the up quark had charge $+2e_p/3$ and the down and strange quarks 
had charge $-e_p/3$ (an antiquark had opposite strangeness, charge and 
intrinsic parity from its respective quark). 

\subsection{Simple Quark Models of Hadron Structure}
\label{Sec:barynaive}

Following the introduction of quarks, the proton was considered to be a 
three-quark combination consisting of two up 
quarks and one down quark. This would produce the requisite electric charge 
of $+1$ and strangeness zero. In a similar vein, the $K^+$ meson with 
strangeness $S=+1$ and charge $+1$ could be composed of an up quark and a 
strange antiquark. Furthermore, the spin and parity of a baryon or meson 
could in principle be constructed by forming various spin-parity combinations 
of the requisite quarks. 

Several groups began to consider what are now known 
as \textit{constituent quark models} to investigate the apparent properties 
of these quark combinations, in terms of the known spins, parities and 
masses of known states \cite{Clo79}. Beginning in the mid 1960s, such efforts 
had some notable successes. Many of the properties of the baryon octet and 
decuplet could be reproduced in a constituent quark model with up and down 
quarks of roughly 340 MeV mass and a strange quark mass in the vicinity of 
450 MeV. The three valence quarks in the proton would occupy the lowest 
states in 
a confining potential such as an harmonic oscillator. The up quark spins 
were parallel to the direction of the proton's spin, and the down quark was 
anti-parallel, with the three quarks coupled to overall spin $1/2$ 
and isospin $1/2$. The $\Delta^{++}$(1232) state was composed of three up 
quarks, in the same confining potential. Here the quark spins
are aligned, producing overall spin-parity $3/2^+$. The strength of the color 
magnetic interaction could be inferred from the $N-\Delta$ mass splitting. 
The $\Lambda$ is composed of $uds$ valence quarks. 
The $N-\Lambda$ splitting gives the mass of the strange constituent 
quark. In this model of the $\Lambda$, the light quarks are coupled to spin 
$0$ and the spin of the strange quark is parallel to the $\Lambda$ spin 
direction. In a similar vein, the lowest nonets of pseudoscalar and 
vector mesons are made by coupling $q\bar{q}$ pairs.   

Despite their apparent success in describing the lightest hadron multiplets, 
constituent quark models encountered a series of problems. 
Some of these were: (1) The constituent quark model predicts that the first 
excited baryons should have negative 
parity. However, the lowest-lying nucleon excited state, the $N(1440)$, has 
positive parity \cite{PDG,Clo79}. Furthermore, the masses, spins and parities 
of excited 
baryonic and mesonic levels disagreed with constituent quark predictions. 
(2) Constituent quark masses consistent with the lowest baryon and meson 
multiplets dramatically over-predicted the pion mass of about 140 
MeV. (3) The constituent quark model suggested that quarks 
in a nucleon were rather lightly bound and non-relativistic. Arguing from 
experience with atomic nuclei, the constituent quark model suggested that 
it should be reasonably easy to eject an individual quark from a proton and 
observe it in a detector. 
(4) This suggested that fractional charges could be emitted from nucleons. 
Many extremely sensitive experiments were mounted to search for fractional 
charges; apart from one or two anomalous results, no isolated fractional 
charges were ever detected \cite{Jon77}.
(5) Constituent quark models required the quarks to be totally symmetric upon 
interchange of space, spin and isospin quantum numbers. Constituent quark 
(fermion) wave functions thus violated the Pauli principle.  

\subsection{Development of Quantum Chromodynamics}
\label{Sec:devQCD}

The non-observation of free quarks and violation of the 
Pauli principle led some to postulate that quarks might not be 
real but might simply represent a mathematical `mnemonic' 
device enabling one to predict meson and baryon properties. 
However, several arguments against the `reality' of quarks were soon overcome. 
First was the discovery of the property of \textit{color}, an additive 
property of quarks which takes on three values \cite{Gre64}. All stable 
particles exist only in combinations of quarks and antiquarks that couple to 
zero net color. For baryons, the resulting three-quark wavefunction is 
completely color-antisymmetric. Thus quark wavefunctions obey the Pauli 
principle when one includes color degrees of freedom.  

The next significant evidence for quarks came from deep 
inelastic scattering (DIS) reactions \cite{Hod97}. These involved scattering 
of leptons 
(charged leptons, or neutrino-induced charge-changing reactions) from 
nucleons at very large momentum transfers; one observed 
only the outgoing lepton and no information regarding the struck 
nucleon(s). Such reactions showed distinctive features 
consistent with scattering from elementary pointlike objects inside the 
proton and neutron. Later in these lectures, we will discuss in considerable 
detail the history of DIS reactions, what they reveal about baryon 
structure, and our current understanding of quark-parton distributions 
in the nucleon. 

The quark model was extended with the discovery of three additional quark 
flavors. In 1974 the charm quark was discovered, with mass roughly 1500 
MeV \cite{Aub74}. In 1977 the bottom quark was discovered, with a mass 
about 4500 MeV \cite{Her77}. Finally 1995 marked the discovery of the top 
quark, with a mass roughly 171,000 MeV \cite{Abe95}. The quarks can be 
grouped into three doublets (u,d) (c,s) and 
(t,b); in each pair the first quark has charge $+2/3$ and the second 
quark $-1/3$. Except for the light quarks, the doublets are also 
characterized by exceptionally large mass splittings. 

With the discovery of heavier quark flavors one expects to see 
additional multiplets of particles containing the heavy quarks 
(the right side of Fig.~\ref{Fig:multiplets} shows low-lying multiplets 
with 4 flavors). We now know 
that the up and down quark masses are much smaller than the `constituent' 
quarks ($u$ and $d$ `current quark' masses are roughly 4 MeV and 9 MeV,
respectively). Thus, the nucleon mass contains very large 
contributions from sea quarks and gluons, and there is no way that a 
non-relativistic perturbative treatment can explain the properties of 
baryons containing only light quarks. On the other hand, charm and bottom 
quarks are sufficiently massive that simple models should accurately 
describe hadrons containing these quarks. Heavy quark effective theory (HQET) 
has made great use of the simplifications that arise due to the very 
large mass of the heavy quarks \cite{Isg89}. 

We have now a detailed theory of quark interactions. This has culminated in 
a non-Abelian gauge theory of quarks interacting 
through the exchange of colored gluons, the carriers of the strong force 
between quarks. This theory is known as \textit{quantum chromodynamics} or 
QCD \cite{Ell96}. A first breakthrough was an understanding of the 
phenomenon of 
\textit{asymptotic freedom}. Through the Nobel Prize-winning work of Gross 
and Wilczek and Politzer \cite{Gro73}, it was 
demonstrated that QCD had the property that at high energies (or 
alternatively, at very short distances) the interaction between quarks became 
progressively weaker and weaker. At these very high energies, one could 
approximate interactions with quarks in terms of free quark interactions. 
Asymptotic freedom demonstrated why deeply inelastic scattering at very high 
momentum transfers looks like scattering from free constituents. 

\begin{figure}[ht]
\includegraphics[width=5.5in]{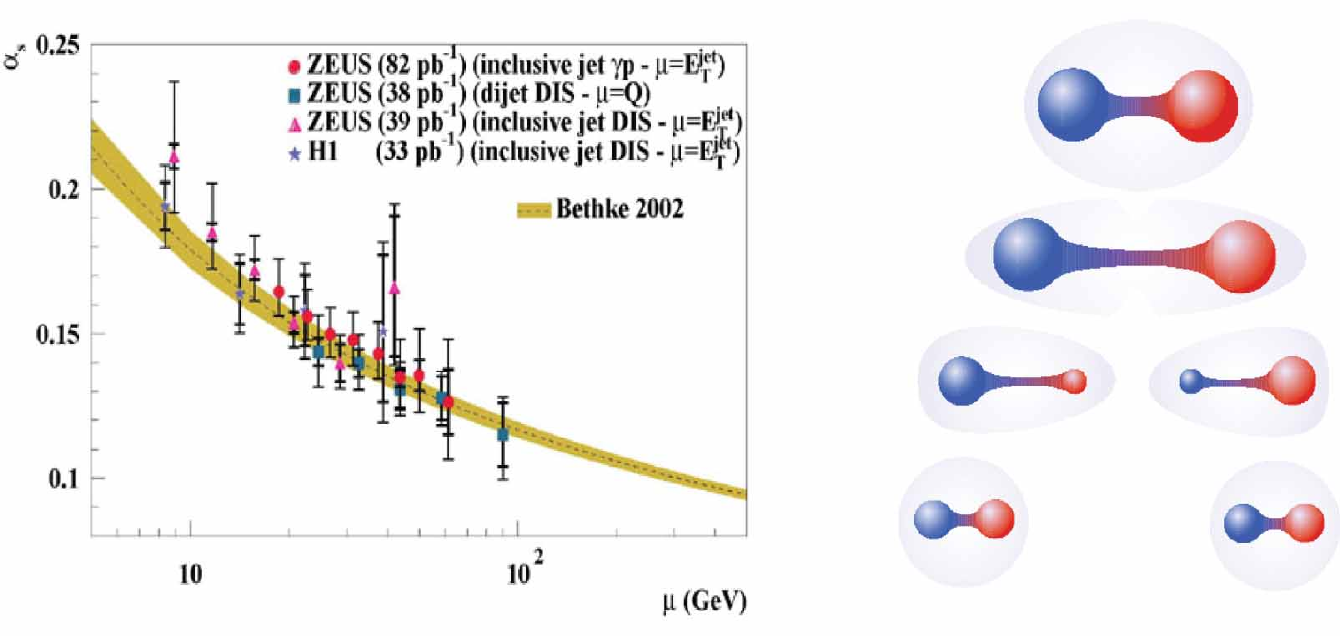}
\caption{\label{Fig:asymptot} Left: the value of the QCD strong coupling 
constant $\alpha_s(E)$ vs.~the energy $E$ in GeV. The decrease of the coupling 
constant with energy demonstrates the phenomenon of asymptotic 
freedom. Right: schematic picture of quark confinement. As quarks are 
separated, a 'string' or flux tube forms. When the string is broken a new 
$q-\bar{q}$ pair is formed.}
\end{figure}

The left-hand graph in Fig.~\ref{Fig:asymptot} shows the strong coupling 
constant $\alpha_s$ as a function of energy. The property of asymptotic 
freedom 
is demonstrated by the monotonic decrease in $\alpha_s$ with increasing 
energy. Conversely, the strong coupling strength increases rapidly at low 
energies. This means that an understanding of QCD at low energies must 
deal with very strongly-interacting systems. This is additionally 
complicated by the realization that gluons have strong self-coupling,  
unlike the situation in QED.  
QCD offered formidable challenges to the development of models of 
strongly interacting systems.  

A second major feature of QCD is that an individual colored 
quark experiences a force that grows monotonically at large 
distances. This is shown schematically in the right-hand picture in 
Fig.~\ref{Fig:asymptot}. As a quark-antiquark pair is separated a string 
or flux tube forms. If one attempts to break the string confining an 
individual quark, eventually one will produce an additional $q-\bar{q}$ pair 
(a meson), rather than liberating the quark. Confinement has never been 
proved analytically in QCD, however it provides a natural explanation for the 
experimental failure to observe individual quarks. Confinement also occurs in 
lattice gauge QCD.  

The QCD Lagrangian has the form  
\bea
\LQCD &=& \overline{q}\left( i\gamma^{\mu}{\cal D}_{\mu} - m \right)
  q - \frac{1}{4}\Gaup\Gadn \nonumber \\ 
  \Gaup &=& \paru\Ganu - \parv\Gamu - g\fabc\Gbmu\Gcnu 
\label{eq:LQCD}
\eea
In Eq.~(\ref{eq:LQCD}) one sums over all flavors  
of quark fields, represented by $q$, the quantities $\Gamu$ represent the 
eight gluon fields, and the term $\Gaup$ is the QCD analog of the 
electromagnetic tensor $\Fmn$ which occurs in quantum electrodynamics. 
An excellent summary of QCD is given in the Handbook of Perturbative 
QCD \cite{Ste95}. 

As we have mentioned, the problem of solving the QCD Lagrangian at low 
energies turned out to be formidable. Our best estimate of $u$ and $d$ 
quark mass is less than 10 MeV, which leads 
to the astonishing conclusion that the `current quark' masses  
make up only about 2\% of the proton mass! The remainder of the mass 
must consist of very large contributions from gluons and the sea. 
At present, the only viable means to solve QCD at low 
energies appears to be \textit{lattice gauge} techniques. In this case 
one discretizes space-time on a lattice. Good numerical results 
can be achieved by rotating to Euclidean (imaginary) time. The 
continuum limit is achieved as the lattice spacing $a \rightarrow 0$. 
One discretizes the Yang-Mills action for the QCD Lagrangian, defines 
the quark fields on the lattice sites, and uses Monte Carlo techniques 
for solving the resulting multi-dimensional integrals. Lectures on 
lattice gauge theory will be given at this workshop by N. Mathur \cite{Mat08}. 

\section{DIS and the Partonic Structure of Baryons}
\label{Sec:partonic}

Our current view is that hadrons are composed of 
structureless elementary particles, quarks and antiquarks, whose 
strong interactions are mediated by exchange of colored gluons. We now 
review the history of the experimental data that demonstrated the quark 
nature of baryon structure, and indicates that quarks have no internal 
structure. The first significant evidence for the elementary nature of 
quarks came from deep inelastic scattering (DIS) reactions \cite{Hod97}. These 
involved scattering of leptons (initially charged leptons, or neutrino-induced 
charge-changing reactions) from nucleons at very large momentum transfers. In 
these reactions one observed only the outgoing charged lepton and no 
information regarding the struck nucleon(s). Such reactions 
carried out initially with electron beams at the Stanford Linear Accelerator 
Center (SLAC) and with neutrinos at CERN, showed distinctive features 
consistent with scattering from elementary pointlike objects inside the 
proton and neutron. Initially these constituents were called `partons' 
but these were later shown to be quarks and antiquarks \cite{Clo79}. 

\begin{figure}[ht]
\includegraphics[width=1.5in,angle=0]{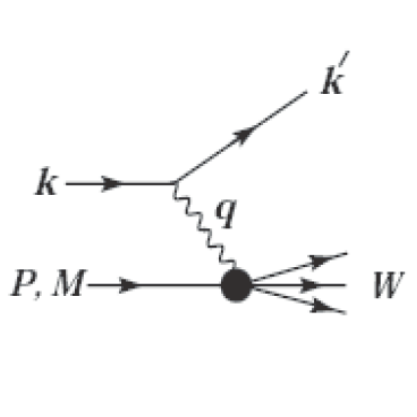}
\caption{\label{Fig:eDIS} Schematic diagram of deep inelastic scattering 
of charged lepton, with initial and final momenta $k$ and $k'$,   
from a nucleon with momentum $P$, through exchange of a virtual photon.}
\end{figure}

Consider DIS reactions induced by scattering of charged leptons from 
a nucleon, mediated by exchange of a virtual photon, as shown schematically 
in Fig.~\ref{Fig:eDIS}. We can describe the process in terms of two 
relativistic invariants, which we will define as $Q^2$ and $x$, defined as 
$Q^2 = -q^2$ and $x = Q^2/(2M\nu)$; here the 4-momentum transfer $Q^2$ is 
large, and $\nu = E - E'$ is the lepton energy loss. 

The cross section for DIS with a charged lepton can be written as the 
contraction 
of a lepton tensor $L^{\mu\nu}$ and a hadron tensor $W_{\mu\nu}$, where 
\bea
\frac{d^2\sigma}{dxdy} &=& \frac{2\pi\alpha^2}{Q^4} \, L^{\mu\nu} W_{\mu\nu} 
  \nonumber \\ W_{\mu\nu} &=& \left(-g_{\mu\nu} + \frac{q_\mu q_\nu}{q^2}
  \right)F_1^\gamma(x,Q^2) + \frac{P_\mu P_\nu}{P\cdot q} F_2^\gamma(x,Q^2)
\label{eq:Fdef}
\eea
The cross section for lepton-nucleon DIS (mediated by photon exchange) thus 
depends upon two structure functions $F_1^\gamma$ and $F_2^\gamma$. Each  
structure function depends on the invariants $x$ and $Q^2$. In 
a frame of reference where the proton is moving with infinite speed, the 
quantity $x$ gives the fraction of the nucleon's momentum carried by 
the struck parton. In Fig.~\ref{Fig:eDIS}, one could also exchange a weak 
vector boson $Z_0$ between the charged lepton and nucleon. This would give 
rise to a third structure function denoted $F_3$, which would 
be associated with parity-violating transitions.  

\begin{figure}[ht]
\includegraphics[width=3.5in,angle=0]{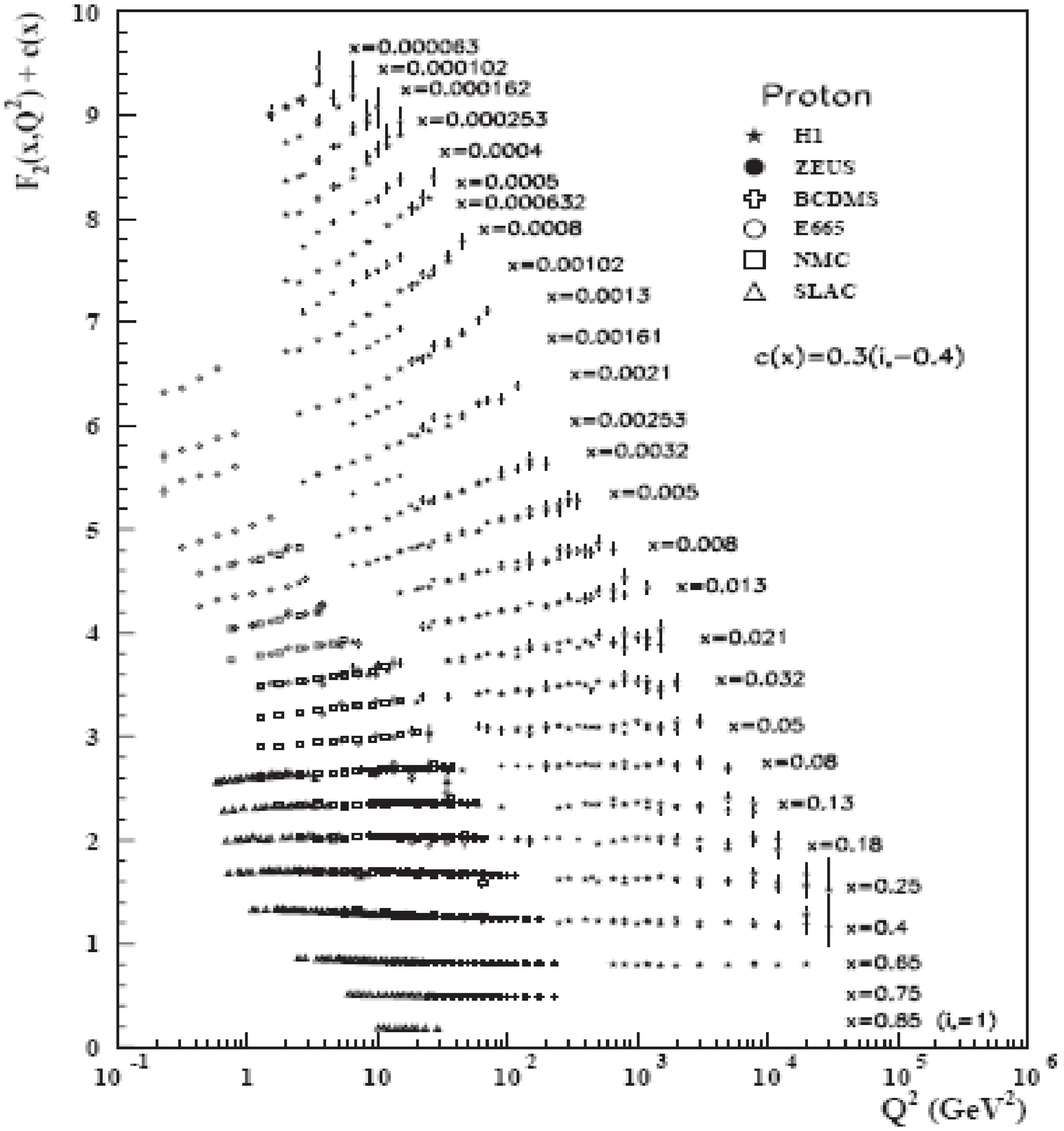} 
\includegraphics[width=2.7in,angle=0]{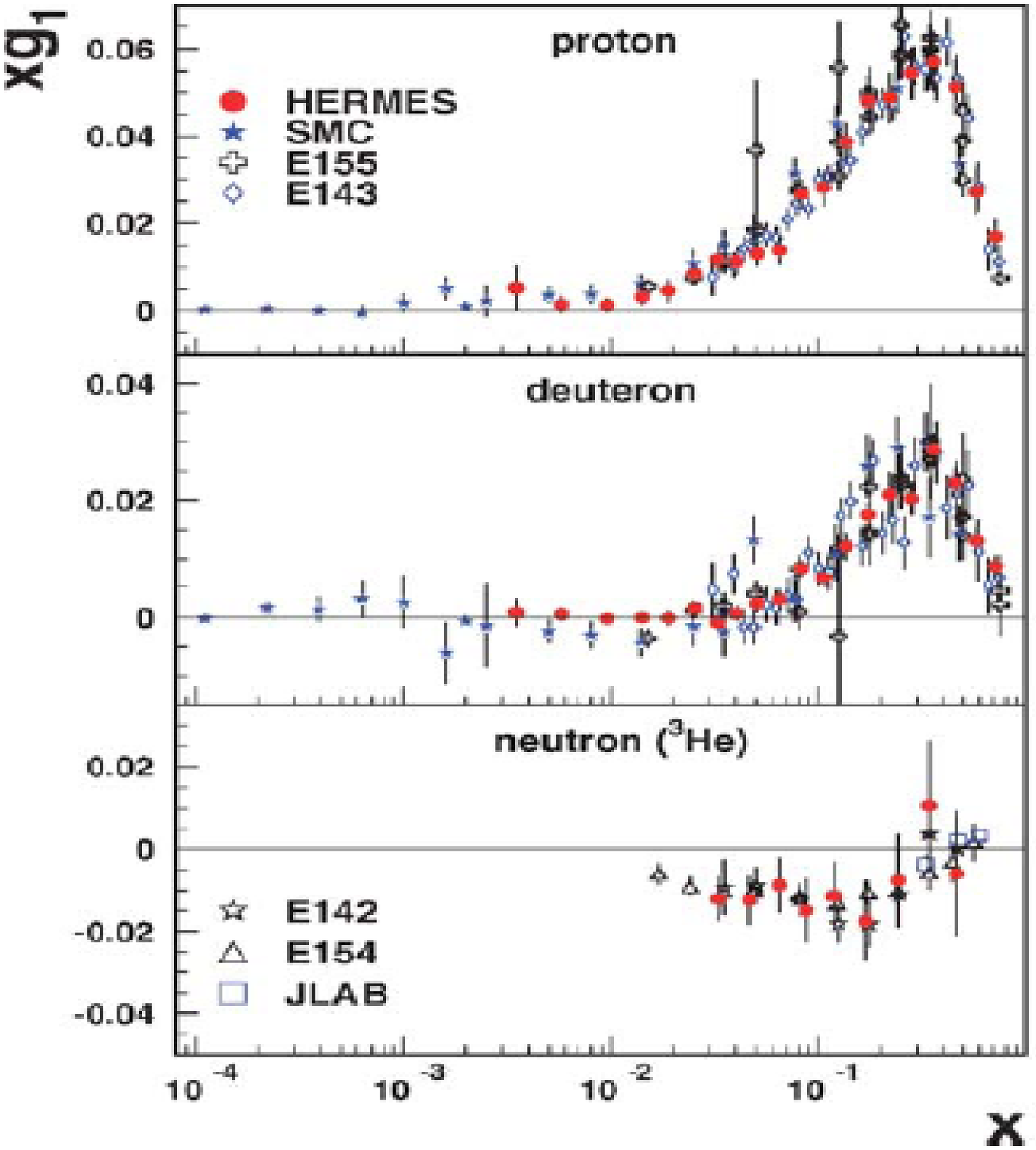} 
\caption{\label{Fig:F2} Left: the structure function $F_2^\gamma$ obtained 
from DIS from charged leptons on protons. $F_2$ is plotted for different 
values of $x$ as a function of $Q^2$. Right: polarized structure function 
$xg_1(x,Q^2)$, obtained for a fixed value $Q^2 = 10$ GeV$^2$ and plotted 
for $p, D$ and $n$.}
\end{figure}

Over the past thirty years, many experiments have been devoted to 
measuring these structure functions. The left-hand side of Fig.~\ref{Fig:F2} 
shows measurements of the structure function $F_2^\gamma$.  
Values of $F_2^\gamma$ are plotted for various values of $x$, vs.~$Q^2$. 
The curves for different $x$ values are offset so that they do not lie 
on top of one another. The values have been obtained from a series of 
different experiments \cite{F2summ}. There are electron-induced experiments 
from the SLAC linear collider and the HERA electron-proton collider 
(its two detectors, H1 and ZEUS), and experiments using 
high-energy muons from the BCDMS and New Muon Collaboration (NMC) 
at CERN and experiment E665 from Fermilab.  

The first dramatic feature of the resulting structure function is the 
precision that has been obtained with these $F_2$ measurements. Except for the 
very highest values of $Q^2$ that can be achieved for a given $x$, the 
errors are very small indeed. The second striking feature is that, over a 
$Q^2$ range of five orders of magnitude, the structure function $F_2^\gamma$ 
varies very little. The constancy of the $F_2$ structure 
functions was clear evidence that scattering was occurring from 
elementary pointlike constituents; otherwise the structure function would 
be governed by a form factor, and over this wide range of $Q^2$ a form factor 
should decrease by orders of magnitude. Friedman, Kendall and 
Taylor shared the 1990 Nobel Prize in physics for their DIS measurements 
in $e-p$ reactions at SLAC that first demonstrated this `scaling' behavior 
in the structure functions \cite{Blo69}. 

One can also measure polarized structure functions. For scattering of 
unpolarized leptons on polarized nucleons, the cross section asymmetry 
resulting from virtual photon exchange is related to the polarized 
structure function $g_1(x,Q^2)$, where one has 
\be
\frac{d^2\sigma (N_\Rightarrow^\rightarrow) - d^2\sigma 
  (N_\Leftarrow^\rightarrow)}{dx\,dy} \sim xg_1(x,Q^2) 
\label{eq:g1def}
\ee
In Eq.~(\ref{eq:g1def}), the cross section 
 $\sigma (N_\Rightarrow^\rightarrow)$ 
is taken for a nucleon whose spin is parallel to the direction of the 
incident lepton beam, and subtracted from the cross section 
$\sigma (N_\Leftarrow^\rightarrow)$ where the nucleon's spin is antiparallel. 
The right-hand side of Fig.~\ref{Fig:F2} plots $xg_1$ vs.~$x$ for experiments 
with protons, deuteron, and neutron (the neutron result is extracted from 
experiments on $^3$He) \cite{G1summ}. Here, the experiments are all 
extrapolated 
to the same value $Q^2 = 10$ GeV$^2$. One can also extract additional 
structure functions when the nucleon spin is transverse to the plane 
of the four-momentum transfer. 

\subsection{A Qualitative Picture of Proton Structure}
\label{Sec:Pqual}

\begin{figure}[ht]
\includegraphics[width=5.0in,angle=0]{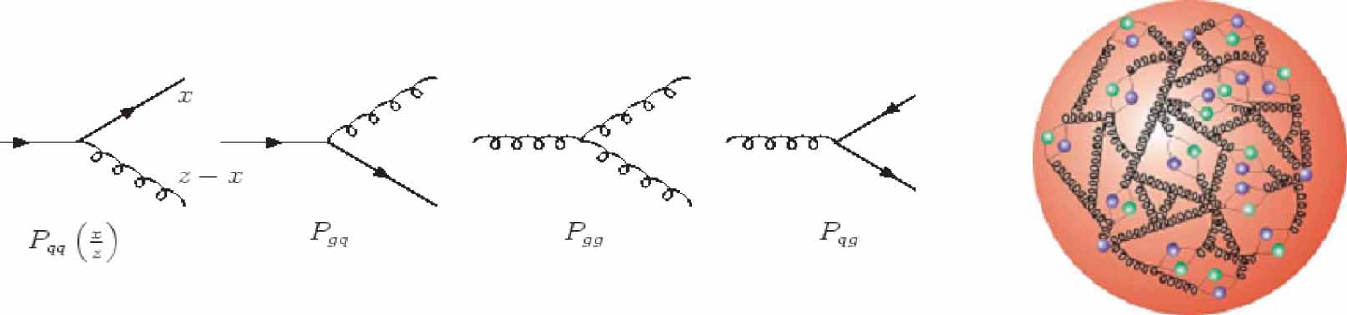}
\caption{\label{Fig:quarkcoup} Left: schematic picture of quark and gluon 
couplings. A quark can radiate a gluon; gluons exhibit self-coupling; and a 
gluon can radiate a quark-antiquark pair. Right: a cartoon showing a 
'valence plus sea and glue' picture of the proton.}
\end{figure}

In Sec.~\ref{Sec:barynaive}, we presented a naive picture of the structure 
of the proton, based on the constituent quark model. Based on our 
current understanding of QCD dynamics, and our knowledge that the `bare' 
masses of $u$ and $d$ quarks are extremely small, we need to modify this 
simplistic picture. We know that at low energies quarks are coupled to gluons
with a coupling constant. Furthermore, 
gluons exhibit self-coupling and are also coupled to quark-antiquark 
pairs. These couplings are shown schematically in the left-hand figure 
in Fig.~\ref{Fig:quarkcoup}. 
 
Imagine that we start with a nucleon containing just three `valence' quarks.  
Through the QCD couplings shown in 
the left-hand side of Fig.~\ref{Fig:quarkcoup}, the valence quarks can radiate 
gluons, which in turn can produce more gluons and quark-antiquark pairs. 
Starting with a proton that contains only `valence' quarks, QCD 
radiation can produce a proton containing a `sea' of gluons and 
quark-antiquark pairs, as depicted by the cartoon on the right in  
Fig.~\ref{Fig:quarkcoup}. 

We have presented two naive pictures of the proton, one with just 
three constituent quarks and the second with many additional 
sea quarks and gluons. What we expect in a given DIS reaction will depend 
upon the value of $x$. Note that from the left-hand side of 
Fig.~\ref{Fig:quarkcoup}, quark or gluon radiation involves sharing 
parton momentum with the additional constituents. 
As QCD radiation produces more and more partons, each parton will 
carry less of the proton momentum and will correspond to 
partons with progressively smaller values of $x$. 
Consequently, the quantity $x$ which denotes the fraction of the proton's 
momentum carried by the struck quark, will determine which picture of 
the nucleon will be `seen' in that reaction. At large values of $x$ 
the nucleon will look like three `valence' quarks; as $x$ decreases 
we expect a `sea' of quark-antiquark pairs and gluons. 

A crucial ingredient is the probability of finding 
partons that carry a given fraction $x$ of the proton's 
momentum. Quantities that satisfy this definition are called {\bf P}arton 
{\bf D}istribution {\bf F}unctions, or PDFs. Define the quantity 
$q(x), \ q = u, \bar{u}, d, \bar{d}, \dots$. For a proton with very high 
momentum, $q(x)$ is the probability that a 
quark of that flavor carries a momentum fraction between $x$ and $x + dx$; 
similarly a gluon PDF $g(x)$ 
gives the probability that a gluon carries momentum fraction $x$. We should 
also include a dependence upon $Q^2$, as we will later show that PDFs have 
a slow (logarithmic) dependence on $Q^2$.  

From Fig.~\ref{Fig:quarkcoup}, antiquarks in the proton will arise mainly 
from gluon radiation of $q-\bar{q}$ pairs. Thus we expect all 
antiquarks to be part of the nucleon `sea' that is radiated by gluons. 
Light ($u$ and $d$) quarks have two sources: one component of these PDFs 
is part of the `valence' distribution; a second component arises  
from gluon radiation. Now $q$ and $\bar{q}$ arising from gluon radiation have 
equal probability. For a given quark flavor we thus define valence quark 
distributions $\qv(x) = q(x) - \bar{q}(x)$. If all antiquark distributions 
arise from gluon radiation, then subtracting the antiquark PDF from the 
quark PDF cancels out the `sea' quark contribution leaving just the `valence' 
part.  

\subsection{Relation Between DIS and Quark PDFs}
\label{Sec:DISPDF}

In a preceding section we showed that DIS reactions could be 
described in terms of a small number of structure functions. An obvious 
question is: what is the relation between the structure functions and the 
quark PDFs that we have just defined? To lowest order in QCD, structure 
functions have very simple relations 
in terms of PDFs. For example, the unpolarized and 
polarized structure functions $F_2^\gamma$ and $g_1^\gamma$ that occur in 
DIS from charged leptons (arising from virtual photon exchange between 
electron and proton as shown in Fig.~\ref{Fig:eDIS}) have the 
form 
\bea
F_2^\gamma (x,Q^2) &=& x\sum_j \, e_j^2 \left[ q_j(x) + \bar{q}_j(x) \right] 
 = x\left[ \frac{4(u + \bar{u})+ d + \bar{d} + s + \bar{s}}{9} \right]   
\nonumber \\ g_1^\gamma(x,Q^2) &=& \sum_j \frac{e_j^2}{2} \left[ 
  \Delta q_j(x) + \Delta \bar{q}_j(x) \right]  \nonumber \\ 
  &=& \frac{4(\Delta u + \Delta \bar{u}) + \Delta d + \Delta \bar{d} + 
  \Delta s + \Delta \bar{s}}{18} \ .
\label{eq:FPDF}
\eea 
In Eq.~(\ref{eq:FPDF}), $e_j$ is the charge of a quark with flavor $q_j$, 
and $\Delta q(x) = \overrightarrow{q}(x) - \overleftarrow{q}(x)$, where 
$\overrightarrow{q}(x)$ ($\overleftarrow{q}(x)$) is the probability 
for finding a quark whose spin is parallel (antiparallel) to the proton spin. 
Eq.~(\ref{eq:FPDF}) includes contributions from three flavors. 

If we include DIS from charged-current $\nu$ leptoproduction, 
\EG~$e^- + p \rightarrow \nu + X$, then the reaction is 
characterized by a virtual $W^-$ being absorbed by the proton. In that 
case three of the structure functions that appear in this reaction 
can be written in lowest order QCD as  
\bea
F_2^{W^-} (x,Q^2) &=& 2x\left[ u + \bar{d} + \bar{s} \right] \nonumber \\ 
 xF_3^{W^-} (x,Q^2) &=& 2x\left[ u - \bar{d} - \bar{s} \right] \nonumber \\ 
  g_1^{W^-} (x, Q^2) &=& \Delta u + \Delta \bar {d} + \Delta \bar{s} \ . 
\label{eq:FgWdef}
\eea
The additional structure function $xF_3^{W^-}$ is present because 
exchange of the weak vector boson $W^-$ contains a piece that violates 
parity; here the structure function $F_3$ characterizes the 
parity-violating part of this interaction. 
Note in both Eqs.~(\ref{eq:FPDF}) and (\ref{eq:FgWdef}), we have not 
included the dependence of the PDFs on $x$ and $Q^2$. 

The above equations show that different observables are sensitive 
to different combinations of PDFs. Thus by carrying out a series  
experiments employing lepton, neutrino and hadron beams, and including  
unpolarized measurements with 
both longitudinal and transversely polarized beams and targets, one 
can gain differential sensitivity to specific quark and gluon PDFs. By 
amassing a large body of experimental data from many different experiments 
one can perform a global analysis of all of the experimental data and extract 
best values for the quark and gluon PDFs \cite{PDG}. 

We mentioned previously that PDFs at different $Q^2$ are related by QCD, 
through what are called the DGLAP evolution equations \cite{Dok77}. These are 
convolution relations relating the PDFs for a given value of $x$ at 
different $Q^2$ values. Thus if one knows the PDFs at one (sufficiently large) 
value of $Q^2$, they can be determined at higher values of $Q^2$ through 
these evolution equations. The non-singlet (valence) PDFs 
satisfy their own integro-differential equation, while the sea quark and 
gluon distributions are coupled, 
\bea
\frac{\partial \qv}{\partial \ln \mu^2} &=& \frac{\als (\mu^2)}{2\pi} 
  P_{qq}\otimes \qv \nonumber \\ 
  \frac{\partial}{\partial \ln \mu^2} 
  \begin{pmatrix} 
   q_s \\ g 
  \end{pmatrix} &=& 
  \frac{\als (\mu^2)}{2\pi}
   \begin{pmatrix} 
   P_{qq} & 2n_f P_{qg} \\ P_{gq} & P_{gg} \end{pmatrix} 
   \otimes \begin{pmatrix} q_s \\ g \end{pmatrix} \nonumber \\ 
 {\rm where} \ \ P \otimes f &\equiv& \int_x^1 \,\frac{dy}{y} \,P(y)\, 
   f\left( \frac{x}{y}\right)
\label{eq:DGLAP}
\eea
In Eq.~(\ref{eq:DGLAP}), the quantity $n_f$ refers to the number of active 
quark flavors participating in the reaction, and the evolution kernels 
$P_{ij}$ can be determined from perturbation theory \cite{Ste95}.

\section{Experimentally Determined Parton Distribution Functions}
\label{Sec:ExPDF}

Over the past three decades a very large number of precision experiments have 
been performed, designed to elucidate the partonic structure of 
nucleons. The pioneering experiments at SLAC were fixed-target 
experiments, with a beam of high-energy electrons incident on a fixed 
target. Such experiments can reach the largest values of $x$, and 
are being continued today at Jefferson Laboratory. Another class 
of experiments involved colliding beams.  For example, the HERA 
accelerator at DESY, which has recently been closed down, was an 
asymmetric collider where 30 GeV electrons collided with 820 GeV 
protons. Fig.~\ref{Fig:F2} shows that the two HERA detectors 
H1 and ZEUS covered an enormous range of $x$ and $Q^2$, and also 
reached to the lowest values of $x$. HERA also had a smaller 
fixed-target program, HERMES, which studied electron collisions 
with polarized nucleon targets. 

Useful information regarding parton distributions has been obtained 
from Drell-Yan processes \cite{Dre70}. These processes involve the production 
of large invariant-mass $\mu^+-\mu^-$ pairs from nucleon-nucleon 
and nucleon-nucleus reactions. These processes occur when a $q$ 
($\bar{q}$) from the projectile annihilates a $\bar{q}$ ($q$) 
of the same flavor from the target, producing a virtual photon 
that decays to a pair of oppositely charged muons. Such processes 
are quite sensitive to the $\bar{q}$ PDFs in the proton. 

Information on $s$ quark distributions is obtained from 
neutrino charge-changing reactions. In such processes an incident 
$\nu_{\mu}$ produces a $\mu^-$ and a virtual $W^+$.  The $W$ 
strikes an $s$ quark and produces a $c$ quark. The charm quark 
subsequently undergoes semi-leptonic decay producing a $\mu^+$. 
The signature of these reactions is production of oppositely charged 
high-energy muons from a neutrino beam, and provides the most 
reliable information on $s$ quark PDFs. Production of $\mu^+-\mu^-$ 
pairs from a $\bar{\nu}$ beam is sensitive to $\bar{s}$ PDFs \cite{Mas07}. 

Information from a wide array of high-energy experiments is placed 
in a global fit to parton distribution functions. From these experiments 
one can extract the structure functions. By now, the relation between 
the structure functions and PDFs is presented at least to next to leading 
order (NLO) in QCD, and sophisticated global fitting routines 
are employed to extract the parton distributions. Two of the best 
known of these are employed by the CTEQ collaboration \cite{CTEQ}, and by the 
MRST group \cite{MRST}. The DGLAP evolution equations of Eq.~(\ref{eq:DGLAP}) 
are used to connect parton distributions at different $Q^2$ values. The 
resulting PDFs are plotted at one particular value of $Q^2$; PDFs at different 
values of $Q^2$ can be obtained from the evolution equations.  

\begin{figure}[ht]
\includegraphics[width=5.5in,angle=0]{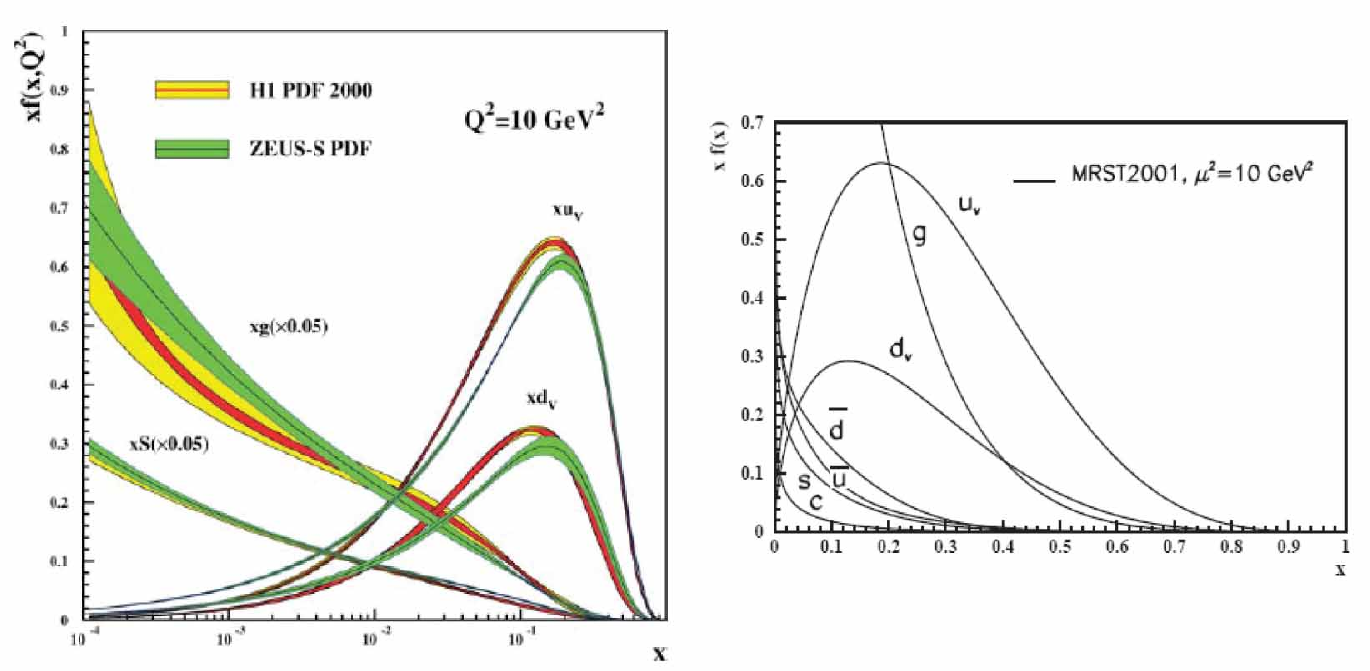}
\caption{\label{Fig:PDFs} Proton PDFs. Values of $xq(x)$ determined from 
global fits to high-energy data evolved to $Q^2 = 10$ GeV$^2$. Left: width of 
band gives a measure of the uncertainty in the PDF. Yellow: determined from 
H1 HERA data; green: determined from ZEUS data. Right: a linear plot in $x$, 
separating different sea quark flavors.}
\end{figure}

Fig.~\ref{Fig:PDFs} shows our current knowledge of quark PDFs. Quark and gluon 
PDFS multiplied by $x$ are plotted vs.~$x$. The left figure is a semi-log plot 
and the right-hand figure is a linear plot that separates different sea quark 
flavors. The PDFs are evolved to $Q^2 = 10$ GeV$^2$. The HERA data is most 
extensive. There are small differences between the PDFs depending on 
data from the H1 detector (yellow band) or from ZEUS (green band) 
\cite{F2summ}. The width 
of the bands gives an indication of the uncertainty in the PDFs. The quantity 
$xS$ is the sum of all sea quark PDFs.    

Our naive picture suggested that at large $x$, 
the proton should be composed primarily of valence quarks. This agrees with 
the analyses; for large $x$ the valence quarks dominate the 
PDFs. For small $x$ the sea and glue distributions rise rapidly. Note 
that in the left graph of Fig.~\ref{Fig:PDFs} the sea and gluon distributions 
have been divided by 20. Thus the sea and glue grow extremely rapidly at 
small $x$.  

The valence PDFs obey quark normalization conditions. Their first moments 
give the total number of valence quarks in the proton,  
$\langle \uv(x)\rangle = 2$, $\langle \dv(x)\rangle = 1$. where the 
brackets denote integration over all $x$. One thus expects that $\uv(x) 
\sim 2\dv(x)$. Fig.~\ref{Fig:PDFs} shows  
that this is approximately the case. 
 
In the right-hand side of Fig.~\ref{Fig:PDFs} the sea PDFs are 
separated by flavor. The $s$ quark distribution is roughly half the average 
of $u$ and $d$ sea quark PDFs. The $c$ quark distribution is much smaller 
than the other sea quark distributions, due to the much larger 
$c$ quark mass. The right side of Fig.~\ref{Fig:PDFs} shows that 
the $\bar{d}$ distribution is larger than the $\bar{u}$ distribution. We can 
understand this from `meson cloud' models of the nucleon \cite{Spe97}. 
Such models include  
contributions particularly from the pion cloud. The proton can undergo virtual 
transitions $p \rightarrow n + \pi^+$. The valence quark content of the 
$\pi^+$ is $u-\bar{d}$. Thus one contribution to the $\bar{d}$ distribution 
arises from scattering from the $\bar{d}$ in the 
$\pi^+$ cloud of the proton. `Meson cloud' models can 
account at least qualitatively for the magnitude and $x$ distribution of 
the $\bar{d} - \bar{u}$ distribution in the proton. 

\section{Surprising Features of Parton Distribution Functions} 
\label{Sec:PDFsurprise}

In the preceding section we reviewed qualitative features of quark PDFs 
obtained from global fits to high energy. Now we will focus on some of 
the surprises that have occurred as we have unraveled the 
structure of the proton. 

\subsection{Momentum carried by the Proton}
\label{Sec:Pmom} 

In our simplest picture with a proton composed of three 
constituent quarks, each `constituent' quark should carry 
roughly $1/3$ of the proton's momentum. The valence quark 
distributions shown in Fig.~\ref{Fig:PDFs} peak at lower values of $x$; 
$\uv$ peaks at $x \sim 0.2$ and $\dv$ peaks at 
an even smaller $x$ value. A major surprise occurred with the 
quantitative understanding of the distribution of the proton's momentum. 
The total proton momentum can be represented by a sum rule  
\be 
 \left\langle \,\sum_j x(q_j(x) + \bar{q}_j(x)) + xg(x) \,\right\rangle = 1
\label{eq:momsum}
\ee
Eq.~(\ref{eq:momsum}) expresses the fact that the proton momentum can be 
divided into the total amount carried respectively by quarks, antiquarks and 
gluons. At high values of $Q^2$, the total proton momentum carried by valence 
quarks is roughly 35\%, sea quarks carry 15\%, and gluons carry 50\% of the 
momentum. This differs 
dramatically from the naive picture where valence quarks would 
carry all of the proton's momentum. It is one more indication of the 
extremely important role of glue, both in constituting the mass of the 
nucleon and now in terms of the proton linear momentum.  

\subsection{Nuclear Modification of Structure Functions}
\label{Sec:nucPDF} 

Another major surprise is the realization that even at very high 
energies, structure functions in nuclei differ from those in the proton. 
First let us discuss what we expect for parton distributions in the 
neutron. The operation of charge symmetry (CS) interchanges $p$ and $n$; this 
is a specific rotation in isospin space (a rotation of $180^\circ$ about 
the `2' axis in isospace) that interchanges $p$ and $n$ labels. At low 
energies, charge symmetry is obeyed extremely 
well, with most low-energy amplitudes obeying CS at the 1\% level or 
better. So, it is `natural' for CS also to hold at 
high energies. In principle one can test the validity of parton 
charge symmetry; no violations have been observed and present upper limits 
of partonic charge symmetry are at the level of several percent \cite{Lon98}. 

If one assumes the validity of charge symmetry at the partonic level, 
then all PDFs for the neutron can be written in terms of those for the 
proton. Charge symmetry predicts that $u^n(x) = d^p(x)$ and 
$d^n(x) = u^p(x)$; it also predicts that $s$ and $c$ PDFs
should be identical for $n$ and $p$. Analogous equations 
hold for antiquarks. 
Since heavier nuclei have roughly equal numbers of neutrons and protons, 
in comparing nuclear structure functions to those in the nucleon it makes 
sense to compare heavier nuclei with the deuteron. 

\begin{figure}[ht]
\includegraphics[width=4.2in,angle=0]{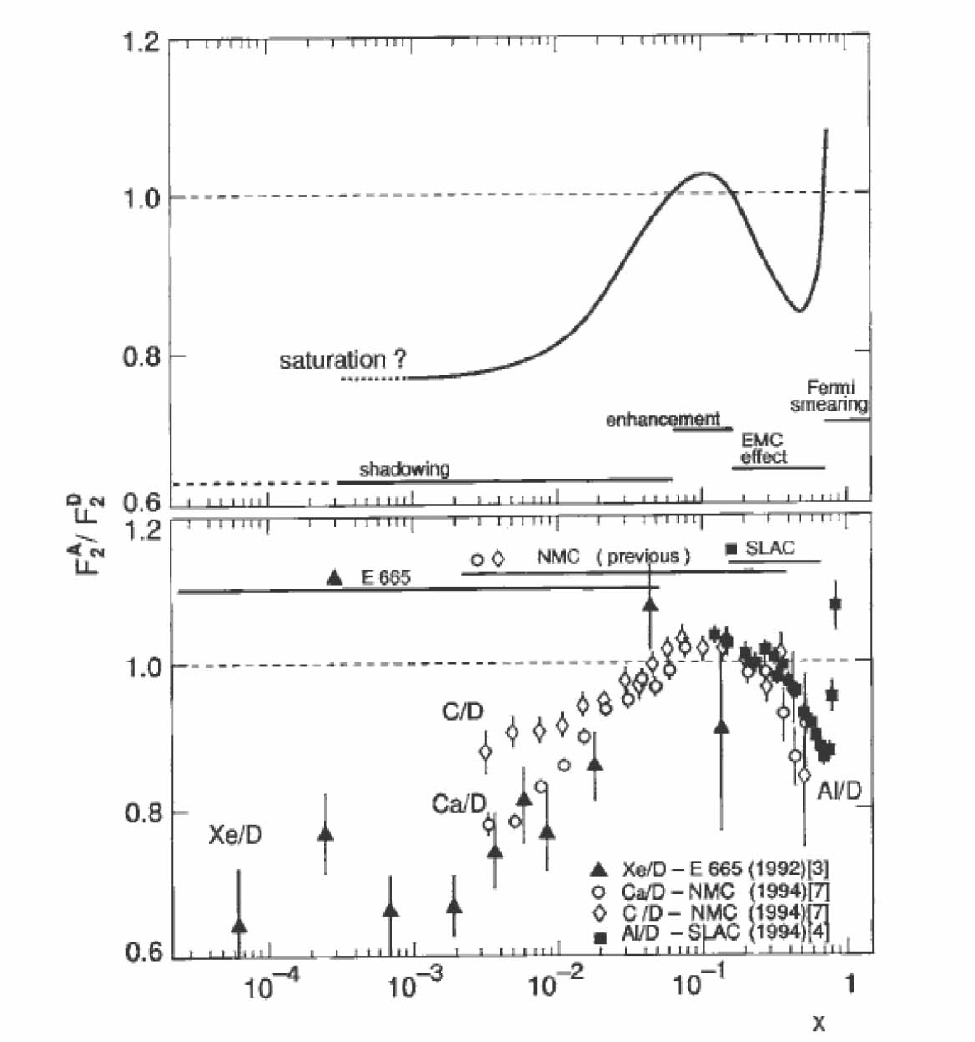}
\caption{\label{Fig:nucrat} The ratio of $F_2$ structure functions for 
nuclei to those for the deuteron, vs.~$x$. Top: a curve that 
shows the qualitative behavior of this ratio vs.~$x$. Bottom: results of 
this ratio for experiments on four nuclei.}
\end{figure}

To date, there have been extensive measurements of the $F_2$ structure 
function for DIS of charged leptons from nuclei.  
One plots the $F_2$ structure function per nucleon arising from virtual photon 
exchange, and constructs the ratio of the 
structure function for a nucleus with $A$ nucleons with that 
for the deuteron,  
\be 
\RAx = F_2^A(x)/F_2^D(x) 
\label{eq:nucrat}
\ee
Fig.~\ref{Fig:nucrat} plots the quantity $\RAx$ vs. $x$ for 
various nuclei, from Arneodo \cite{Arn94}. The top curve shows the 
qualitative behavior of 
$\RAx$ vs.~$x$. The bottom points are the results of this ratio for 
experiments on four nuclei. The diamonds and open circles are results on C 
and Ca respectively from NMC. The solid squares show results on Al from 
SLAC. The solid triangles are results on Xe from E665. 

The experimental points show rather dramatically that the respective structure 
functions $F_2$ have a distinct $A$ dependence. Furthermore, the nuclear 
effects on the $F_2$ structure functions can be divided into roughly 
three distinct regions: a `shadowing' region for small $x < 0.05$; 
a `Fermi smearing' region for very large $x$; and what is called  
the `EMC effect' in the region of $x$ between the shadowing and Fermi 
smearing regions. Fig.~\ref{Fig:nucrat} shows dramatically that even 
at high energies, there remain significant differences between DIS 
from nuclei and scattering from individual free nucleons. 

The $A$-dependence of $\RAx$ for intermediate 
$x$ is called the `EMC effect' after the definitive 
experiments in the early 1980s from the European Muon 
Collaboration at CERN \cite{Gee95}. These experimental results were  
extremely surprising. At least in the regime of intermediate $x$, it was 
widely assumed that at sufficiently high energies, nuclear DIS  
would look just like DIS from quarks in individual free nucleons.  

Since $x = Q^2/(2M\nu)$, the very large $x$ region corresponds to the 
largest values of $Q^2$ for a given incident energy. On a nuclear target, 
scattering of a virtual photon at large $x$ values will be greatly affected 
by the nuclear Fermi momentum. For collisions at very large $Q^2$ 
to the nucleus, nucleons with Fermi momentum moving towards (away from) the 
virtual photon will experience collisions at a relatively much larger 
(smaller) value of $Q^2$. Since $F_2 \rightarrow 0$ rapidly at large $x$, 
Fermi smearing effects will produce nuclear $F_2$ that are much larger than 
that for a free nucleon. Thus we expect to see 
a very rapid increase in $\RAx$ at sufficiently high $x$ due to these 
Fermi smearing effects, in agreement with the results  
shown in Fig.~\ref{Fig:nucrat}.    

\subsubsection{Nuclear Effects in the Shadowing Regime}
\label{Sec:Shad}

Fig.~\ref{Fig:nucrat} shows that in the region of small $x \le 0.05$, 
the ratio $\RAx$ decreases monotonically with decreasing $x$, and  
$\RAx$ shows some sign of approaching a constant value at very small $x$. The 
experimental points show that this curve is not universal but has a 
characteristic $A$ dependence, with the `shadowing' corrections becoming 
larger with increasing $A$. For a given incident energy the small $x$ region 
corresponds to the smallest values of $Q^2$. Even for very high 
energies, sufficiently small $x$ values can correspond to  
the range $1 \le Q^2 \le 5$ GeV$^2$. In this region, viewed from the rest 
frame of the nucleus, we expect to see 
effects arising from transitions of the virtual photon to 
vector mesons. From the CVC (Conserved Vector Current) hypothesis, a 
virtual photon will occasionally make a transition to a vector meson, 
particularly a $\rho(770)$. Since photons interact weakly with matter we 
expect the impulse approximation to be valid, in which case $\RAx$ should be 
one. However, the $\rho$ interacts very strongly and at these 
energies will be strongly absorbed. To a good approximation the cross section 
for the $\rho$ should vary as the area of the nucleus,
$\sigma_{\rho} \sim R^2 \sim A^{2/3}$. These are called `shadowing' 
corrections. As soon as a $\rho$ encounters a nucleon it is absorbed.  
Nucleons on the `far' side of the nucleus are `shadowed' from seeing 
the $\rho$ by the nucleons on the `near' side. 

The nuclear $F_2$ structure function per nucleon arising from 
$\gamma^* \rightarrow \rho$ transitions should behave like $A^{2/3}/A 
\sim A^{-1/3}$. One would expect to see $\RAx \sim A^{-1/3}$ if 100\% of the 
nuclear cross section arose from such transitions. Although quantitative 
calculations are significantly more complicated than this simple picture, 
the qualitative behavior is correct. First, shadowing corrections decrease 
$F_2^A$ in heavier nuclei and one expects 
$\RAx < 1$; the shadowing corrections are largest at the smallest 
values of $x$; and finally shadowing corrections 
become larger with increasing $A$. All of these are observed in the data. 
If we consider nuclear DIS reactions in the infinite momentum frame, then 
shadowing should occur through gluon recombination and interference effects. 
A picture due to Gribov \cite{AGK74} relates shadowing to 
diffractive effects. This picture relates two different scattering phenomena 
and leads to interesting predictions but it is not yet clear that it 
provides quantitative agreement with experiment.  

\subsubsection{The `EMC Effect'}
\label{Sec:EMC}

The final nuclear effect involves a slight increase $\RAx > 1$ just below 
$x = 0.1$, followed by a monotonic decrease 
in the region $0.1 \le x \le 0.6$; this is called the `EMC 
Effect' from its discovery in the early 1980s by the European Muon 
Collaboration. Until that time one expected that nuclear DIS in this region 
would just be the sum of individual lepton-nucleon DIS;  
thus we would expect $F_2(x)$ per nucleon to be the same regardless of $A$.   

\begin{figure}[ht]
\includegraphics[width=5.5in]{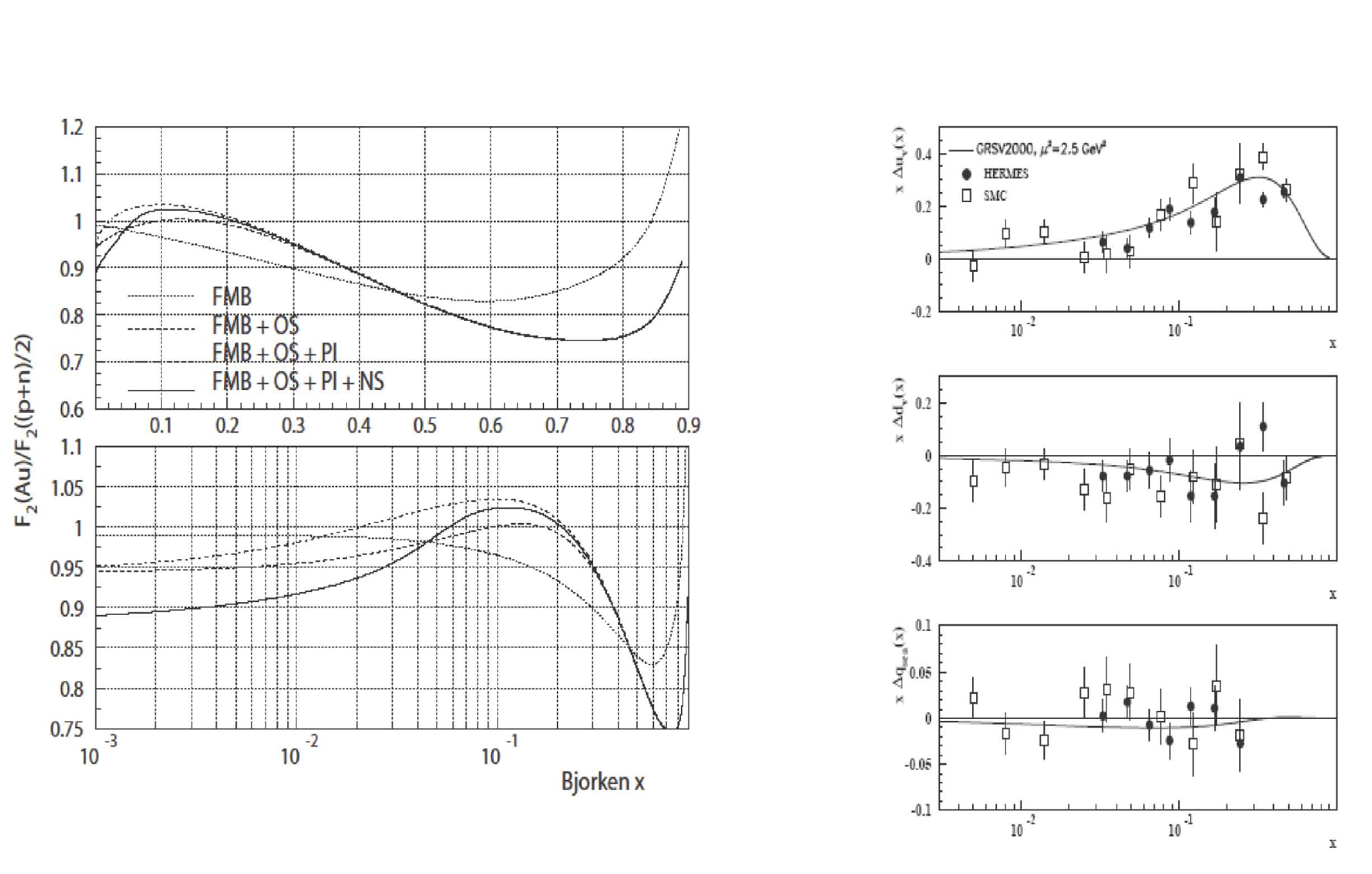}
\caption{\label{Fig:Kulagin} Left: the ratio $\RAx$ from 
Eq.~(\ref{eq:nucrat}) of $F_2$ structure functions for 
gold to those for the deuteron, vs. $x$. Top: linear plot of $\RAx$; 
bottom: semi-log plot. Right: spin-dependent PDFs. Top: $x\Delta\uv (x)$ vs. 
$x$; middle: $x\Delta\dv (x)$; bottom: $x\Delta q_{sea}(x)$. Curves evolved 
to $Q^2 = 2.5$ GeV$^2$.}
\end{figure}

The EMC Effect appears to be nearly universal; the curves 
of $\RAx$ for various nuclei 
lie almost on top of one another. The EMC Effect inspired more than 
1000 papers that considered the origin and explanation of this 
nuclear effect \cite{Gee95}. Kulagin and Petti \cite{Kul06} claim that three 
different mechanisms play a role in 
the EMC Effect. The first two involve nuclear binding and 
off-shell effects on nuclear structure functions. Treating bound nucleons 
non-relativistically and using weak-binding approximations, 
one can write $F_2^A$ as a product of free $F_2$ structure functions for 
$p$ and $n$, convoluted with  
the virtuality of the bound nucleons. An additional contribution to 
$F_2^A$ comes from scattering of the virtual photon from the meson cloud, 
particularly the pion cloud. The left side of Fig.~\ref{Fig:Kulagin} 
plots the ratio $\RAx$ vs.~$x$ for gold \cite{Kul06}. The dotted curve  
includes only Fermi momentum plus nuclear binding (FMB) effects. The dashed 
curve includes the FMB plus off-shell (OS) effects. The dot-dashed curve 
adds nuclear pion (PI) effects; and the solid curve is the 
full result including nuclear shadowing (NS) effects. Shadowing is 
significant only in the shadowing regime. Nuclear pion effects are significant 
only for $x \le 0.2$. In the EMC region the most important processes are 
claimed to be binding and off-shell effects, although others have argued 
that nuclear pion effects are dominant \cite{Gee95}. 

\subsection{Polarized DIS and the Spin of the Proton}
\label{Sec:PDIS}

As was mentioned in Sec.~\ref{Sec:DISPDF}, one can extract the quark 
longitudinal spin-dependent PDFs $x\Dux$ and $x\Ddx$ through measurements 
using polarized protons. Eq.~(\ref{eq:g1def}) relates the polarized structure 
function $g_1(x)$ to the asymmetry in deep inelastic scattering for 
unpolarized leptons on longitudinally polarized nucleons. 
Eq.~(\ref{eq:FPDF}) relates the polarized structure function to the 
spin-dependent PDFs. On the right in Fig.~\ref{Fig:Kulagin} we plot the 
longitudinal spin-dependent PDFs that have been extracted from 
experiment. 

The experimental results \cite{SMC} are from the Spin Muon Collaboration (SMC) 
(open squares) and HERMES (solid circles). HERMES 
took semi-inclusive DIS measurements from polarized proton or deuteron 
targets, where one measured a final-state meson ($\pi$ or $K$) in coincidence 
with the scattered lepton. Such semi-inclusive DIS experiments allow one to 
enhance contributions from particular spin-dependent quark flavors. 
As can be seen from Fig.~\ref{Fig:Kulagin}, one obtains $\Delta \uv(x) > 0$ 
and $\Delta \dv(x) < 0$, and in magnitude $\Delta \uv(x) > |\Delta \dv(x)|$, 
in agreement with naive quark model expectation. It is also clear that 
the sea quark contribution to the nucleon spin is small and consistent 
with zero for all measured $x$ values. 

\subsubsection{The 'Proton Spin Crisis'}
\label{Sec:crisis}

We can write a sum rule for the total spin of the proton, 
\be
\frac{1}{2} = \frac{1}{2}\DS + \DG + L_q + L_g \ \ , 
\label{eq:psum}
\ee
In Eq.~(\ref{eq:psum}), $\DS$ is the total spin carried by quarks and 
antiquarks, $\DG$ is the total spin carried by gluons, and the quantities 
$L_q$ and $L_g$ are respectively the total orbital angular momentum 
carried by quarks and gluons. The capital letters refer to the first 
moment over $x$, \IE 
\be
\DS = \langle \sum_q \Delta q(x) + \Delta\bar{q}(x) \rangle \ ; 
  \hspace{0.4cm} \DG = \langle \Delta g(x) \rangle 
\label{eq:totspin}
\ee  
In the constituent quark picture the 
first moments of quark spin PDFs are given by  
\be
\Delta U = \frac{4}{3}, \,\, \Delta D = -\frac{1}{3}, 
  \,\, \Delta S = 0, \,\,\DS = \Delta U + \Delta D + \Delta S = 1;  
\label{eq:spinconst}
\ee
Eq.~(\ref{eq:spinconst}) shows that in the naive quark picture all  
the proton spin arises from the quarks, with zero contribution 
from gluons or orbital angular momentum. 

The first quantitative measurements of the quark spin PDFs gave 
results like $\DS \sim 0.2 \pm 0.2$. Not only was the total proton spin 
carried by quarks far smaller than 100\%, these results were consistent 
with \textit{zero} proton spin being carried by quarks! This 
surprising and puzzling result was termed the `proton spin crisis.' 
We have now obtained more precise 
results and the current best value is $\DS \sim 0.3 \pm 0.05$. 
So the total proton spin carried by quarks is somewhat larger than it 
had appeared, and it is no longer consistent with zero. Nevertheless, 
one very important quantitative question is exactly how do we account 
for the spin carried by the proton \cite{Ans97}? A related question is whether 
or not the current experimental result $\DS \sim 0.3$ for the proton 
spin carried by quarks is surprising. 

\subsubsection{Contribution of Polarized Glue to Proton Spin}
\label{Sec:gluepol} 

Previously we saw that gluons carried a 
surprisingly large fraction of the proton momentum. It seems natural  
to ask whether gluons might also carry much of 
the proton spin. In order to answer this important question, dedicated 
experiments to measure gluon contributions to the proton spin have been 
mounted at three laboratories. There is the COMPASS experiment at CERN, 
experiments from the RHIC Spin group at the Relativistic Heavy Ion Collider 
(RHIC) at Brookhaven, and the HERMES experiment at HERA. We will discuss 
particularly the COMPASS experiment. 

\begin{figure}[ht]
\includegraphics[width=5.5in,angle=0]{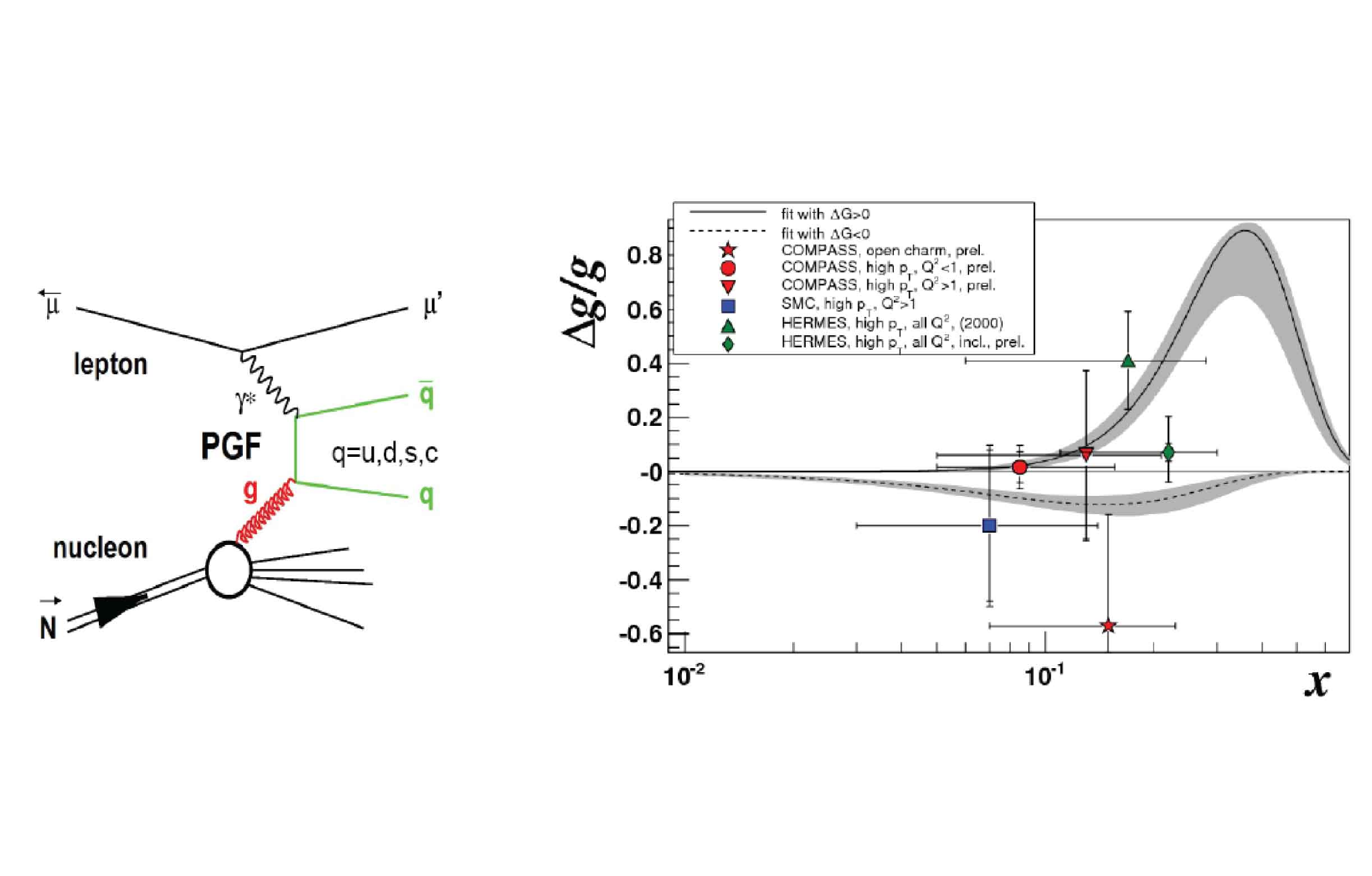}
\caption{\label{Fig:PGF} Left: schematic diagram illustrating  
'photon-gluon fusion.' The virtual photon from a polarized muon couples to 
the gluon distribution through production of a quark-antiquark pair.
Right: $\Delta g(x)/g(x)$ vs.~$x$. Star: COMPASS open-charm measurement; 
circle and downward-facing 
triangle: COMPASS high-$\pT$ measurements with different $Q^2$ cuts; 
square: SMC high-$\pT$ measurement; upward-facing triangle and 
diamond: two different HERMES high-$\pT$ measurements.}
\end{figure}

At COMPASS, the goal was to measure gluon contributions to the proton 
spin through photon-gluon fusion \cite{Bra05}. In this process, shown 
schematically 
in Fig.~\ref{Fig:PGF}, a polarized muon beam couples to a virtual 
photon. This in turn is coupled to the polarized gluon distribution 
in the proton through production of a quark-antiquark pair. The final 
scattered muon is observed in coincidence with a final hadronic state 
which has strong overlap with the $q-\bar{q}$ pair. The COMPASS 
experiment focused on two different hadronic states; the first was 
`open charm,' observation of a final-state hadron containing a charm quark. 
The second situation involved two final-state hadrons with large 
transverse momentum $\pT$. 
  
The right-hand side of Fig.~\ref{Fig:PGF} shows measurements of  
$\Delta g(x)/g(x)$ from several laboratories \cite{Delg}. The star represents 
the COMPASS open-charm measurement. The circle and downward-facing triangle 
are COMPASS high-$\pT$ measurements corresponding to different $Q^2$ cuts. 
The square is a high-$\pT$ measurement from the Spin Muon Collaboration 
(SMC) group. The upward-facing triangle and the diamond are two different 
HERMES high-$\pT$ measurements. The data is all for values $0.05 < x 
\le 0.2$. The values of $\Delta g(x)/g(x)$ are all small and in most 
cases the errors in both $\Delta g$ and the average value of $x$ are 
relatively large. The errors are sufficiently large that the sign of 
$\Delta g$  is not clear. The two curves show fits to these data with 
$\DG >0$ and $\DG < 0$. 

If one takes the COMPASS values one obtains for 
the total gluon contribution to the proton spin, $\DG/G = -0.57 \pm 0.41\ 
(stat) \pm 0.17\ (syst)$ for the open charm measurements, and 
$\DG/G = 0.016 \pm 0.058\ (stat) \pm 0.0.055\ (syst)$ for the high-$\pT$ 
measurements \cite{Age07}. Combining the COMPASS measurements with the 
SMC, HERMES and RHIC spin results, the best value of $\DG$ is small; 
however, it is still statistically possible that $\DG$ could be large.  
Fig.~\ref{Fig:PGF} does not show RHIC spin results that give 
comparable values for $\DG/G$ in a similar $x$ region. 

\subsubsection{Current Status of the Proton Spin Question}\label{Sec:Pspin}

At the present time, despite the large errors on the gluon contribution 
to the proton's spin, it appears that polarized gluons are unlikely to carry 
the majority of the proton's spin. We are left with the question of 
just what fraction of the spin of the proton is carried by polarized 
glue, or by quark or gluon orbital momentum. An area that is being pursued 
at the moment is hard exclusive processes, such as deeply virtual Compton 
scattering or DVCS (where a charged lepton interacting with a nucleon 
exchanges a virtual photon with high $Q^2$, producing a real photon in the 
final state), or exclusive meson production in lepton-nucleon scattering. 
These processes can be described in terms of so-called generalized parton 
distributions or GPDs. It was shown by Ji that a particular moment of the GPDs 
could be related to the total orbital angular momentum of the 
quarks \cite{Ji98}. 

Investigations of DVCS have been undertaken at HERA and at Jefferson 
Laboratory. A major focus of the 12 GeV upgrade at Jefferson Lab will be 
studies of DVCS and exclusive meson production. The physics of the 12 GeV 
upgrade will be reviewed in detail in the first lecture by Tony Thomas at 
this workshop \cite{Tho08}. Thomas' second lecture will summarize his recent 
work with Myrher on the spin of the proton, which will address the 
questions raised in this section.  

\section{New and Proposed Facilities to Study Nucleon Structure}
\label{Sec:newfac}

To date, much information on the partonic structure of the nucleon has 
been obtained from a series of high-energy machines. The first 
experiments that showed scaling in high-energy deep inelastic scattering 
were carried out at SLAC. Following these measurements were continuing 
fixed-target experiments at SLAC, and both fixed-target and collider 
studies at CERN and Fermilab. The HERA asymmetric electron-proton 
collider proved to be a nearly ideal facility to map out parton 
distributions over a very wide range of $x$ and $Q^2$. Neutrino and 
antineutrino beams at Fermilab and CERN allowed one to separate out 
contributions from different quark flavors. Studies of Drell-Yan processes 
at Fermilab and CERN provide processes sensitive to sea quark distributions 
in the nucleon. 

Now that HERA has ceased operations, experiments at CERN and Fermilab 
continue. Various new facilities or upgrades of existing ones have 
the capability of extending our knowledge regarding the partonic 
structure of the nucleon. The first is the Large Hadron Collider or LHC 
which will very shortly begin operation at CERN. The primary thrust of 
LHC operations will be first to find the Higgs particle, a new massive 
particle which is predicted to appear as the result of the spontaneous 
breaking of electroweak symmetry. Next, the LHC will focus on discovering 
either supersymmetric particles or other new phenomena beyond the Standard 
Model. However, the LHC also could have potential applications in determining 
the partonic content of hadrons. 

There are also several new facilities, or upgrades of existing facilities, 
which could examine the partonic content of hadrons. Here I will discuss one 
upgrade of an existing facility and one proposed 
new facility in this regard. The first, the upgrade of the electron 
accelerator at Thomas Jefferson Laboratory, is the subject of the 
first lecture by Tony Thomas at this workshop \cite{Tho08}. As a result, 
I will provide 
only a brief summary of this project; I refer the reader to his talk on 
this subject. The second is a proposed new electron-ion 
collider. Two possible versions of this machine have been suggested, 
one which would be located at Brookhaven National Laboratory and a  
second which would be the subject of a future upgrade at Jefferson 
Laboratory. 

I will very briefly discuss the 
12 GeV upgrade of the CEBAF accelerator at Thomas Jefferson National 
Laboratory \cite{JLAB}. The upgrade will provide 12 GeV continuous electron 
beams to a new Hall D experimental area, which will produce real photon beams 
that should provide precision spectroscopy. This should help to study 
possible `exotic' states, particularly excited mesons that contain one or 
more gluons. In addition, the upgrade would provide up to 11 GeV beams 
to the existing halls A, B and C. 

As was mentioned in the preceding section, a major focus of effort 
following the Jefferson Lab upgrade would be on hard exclusive processes 
such as DVCS or exclusive meson production. It has been shown that 
these hard exclusive processes can be described in terms of integrals 
over quark distributions. These are termed `generalized parton distributions' 
or GPDs. GPDs can provide information regarding the longitudinal momentum 
and transverse position of quarks in the nucleon, and could test the Ji 
sum rule \cite{Ji98}. Following the upgrade, Jefferson Lab will 
have unique kinematic capabilities. It will be able to access parton 
distributions in regions of $x$ and $Q^2$ that complement those 
accessible at laboratories like CERN or HERA. In particular, after the 
upgrade Jefferson Lab will be able to reach regions of Bjorken $x \sim 0.6$ 
and values $Q^2 \sim 8$ GeV$^2$. 

\begin{figure}[ht]
\includegraphics[width=5.5in]{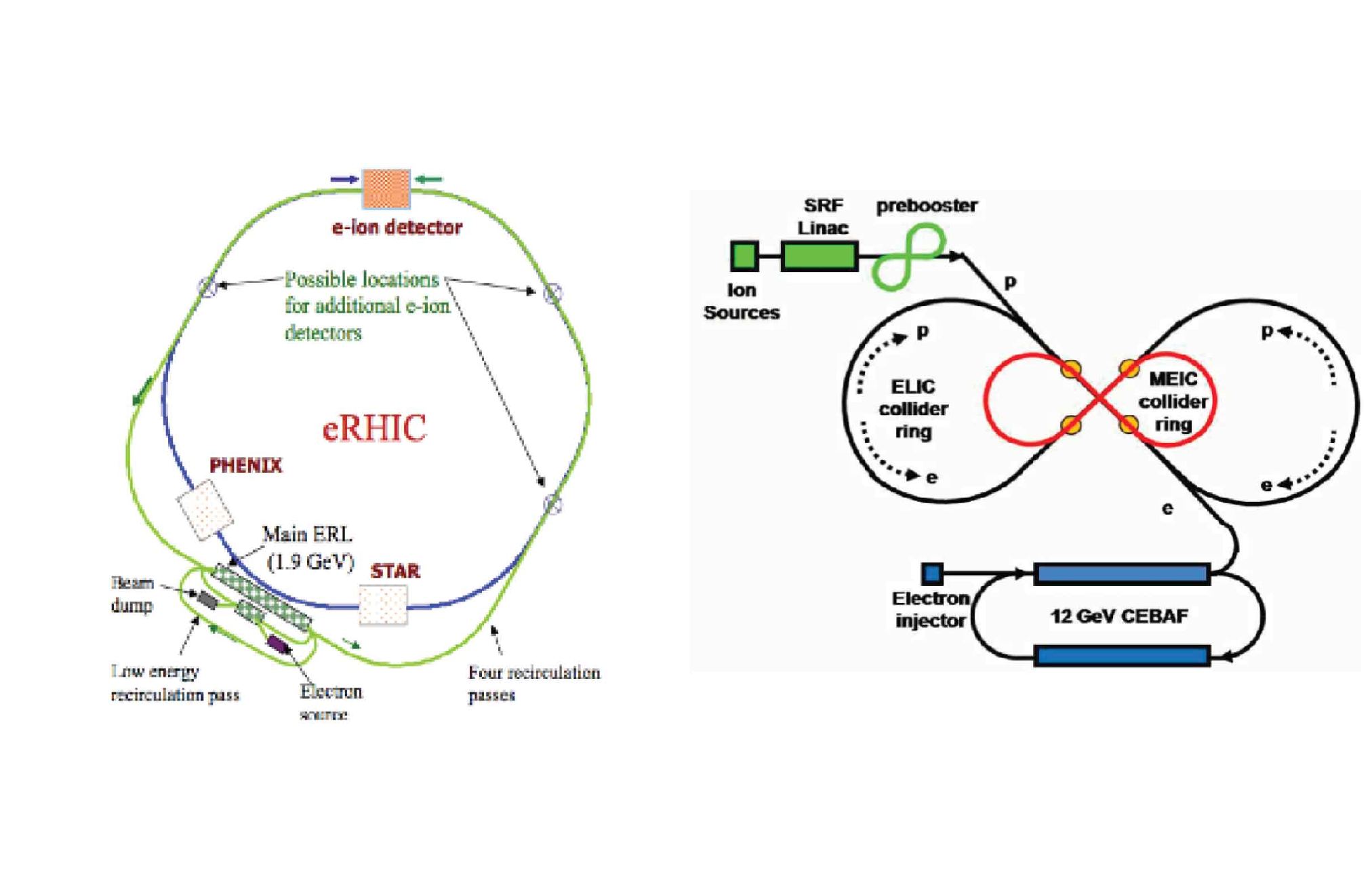}
\caption{\label{Fig:eRHIC} Conceptions of possible 
electron-ion colliders. Left: eRHIC. An electron ring added to the RHIC 
complex at BNL. Right: electron-light ion collider 
at Jefferson Lab. A hadron ring would be added to the CEBAF 
accelerator, accelerating protons and light ions.}
\end{figure}

A proposed new facility is an electron-ion collider. This would consist 
of an electron beam colliding with either light ions or heavy ions. Plans 
to date call for achieving a value $s = 20 - 100$ GeV$^2$, with high 
luminosity. There are two suggestions for such a facility \cite{Des05}. 
In each case this 
would involve building a new accelerator to complement an existing 
machine. The first case would involve building an electron ring at Brookhaven, 
to collide with hadrons from the current RHIC facility. The left-hand 
side of Fig.~\ref{Fig:eRHIC} shows a schematic picture of such a facility. 
A proposed electron ring would be added to the RHIC complex at Brookhaven; 
the ring would provide electrons with energies in the range  5-10 GeV, with 
the possibility of a recirculating linac injector. The electrons would 
collide with beams of heavy or light ions from the RHIC accelerator.

The second scenario would involve building a light ion accelerator at 
Jefferson National Laboratory. This would produce hadrons of 50-100 GeV, 
to collide with electrons of roughly 5 GeV from CEBAF. The right-hand side 
of Fig.~\ref{Fig:eRHIC} shows a schematic picture of such a facility. 
A hadron ring (in this diagram, shown in a figure-eight configuration) 
would be added to the CEBAF electron 
accelerator at Jefferson Lab. The ring would provide protons and light 
ions with energies in the range 50-100 GeV. The proposed hadron accelerator 
would feature four intersection points. Present estimates suggest that very 
high luminosities (${\cal L} \sim 10^{34}/$cm$^2\cdot$sec) might be achieved 
in such a collider.

With an electron-ion collider one could provide quantitative measurements 
of quark spins through semi-inclusive DIS. It would have the possibility 
of measuring PDFs (and particularly gluon distributions) at very small 
Bjorken $x$. Gluon probabilities in the nucleon increase extremely rapidly 
at very small $x$ values, as can be seen from Fig.~\ref{Fig:PDFs}.  
The large nucleon densities in the center of the nucleus could in 
principle support extremely large gluon densities. If gluon 
distributions continue to increase in the interior of the nucleus, at some 
point nuclear gluon probabilities 
would exceed the unitarity bound. Thus one expects at some point for 
many-body effects to slow down this increase. Some have suggested that 
this will lead to a qualitatively new cooperative effect termed a 
'color-glass condensate' \cite{Mcl94}. An electron-ion collider, particularly 
eRHIC which supports heavy-ion hadronic beams, would be able to probe this 
dynamical region. 

An electron-ion collider would also have the capabilities of exploring 
generalized parton distributions over a fairly wide kinematic region. 
Such colliders could also provide detailed quantitative studies of the 
structure of mesons, particularly pions. One other area that could be 
studied with an electron-ion collider is the process of fragmentation. 
This is the process whereby a quark produces a final hadron. Possible 
kinematics for electron-ion colliders would have the capability of 
exploring both beam fragmentation and the so-called target fragmentation 
regions.

The author was supported by NSF contract PHY-0555232, and he 
acknowledges useful discussions with A.W. Thomas and A.P. Szczepaniak.


\begin{thebibliography}{99}

 \bibitem{Clo02} A review of accelerator and detector development with many 
  fine images is Frank Close, Michael Marten and Christine Sutton, {\it The 
  Particle Odyssey}, (Oxford University Press, Oxford 2002). 
 
\bibitem{PDG} W-M Yao \EA, (Particle Data Group) J.~Phys.~{\bf G33} (2006) 1. 

 \bibitem{Gol64} M. Goldberger and K.M. Watson, {\it Collision Theory}, 
 (Wiley and Sons, New York 1964). 
 
 \bibitem{Bre37} G. Breit and E. Wigner, Phys.~Rev.~{\bf 51} (1937) 593. 

 \bibitem{Bog02} M. Boglione and M.R. Pennington, Phys.~Rev.~{\bf D65} 
   (2002) 114010. 

 \bibitem{Dal54} R.H. Dalitz, Phys.~Rev.~{\bf 94} (1954) 1046. 

\bibitem{Ams02} C. Amsler \EA, Eur.Phys.J.~{\bf C23} (2002) 29. 
  
 \bibitem{Gel64} M. Gell-Mann and Y.~Ne'eman, {\it The Eightfold Way}, 
 (W.A. Benjamin, New York 1964). 
 
 \bibitem{Sch63} P.E. Schlein \EA, Phys.~Rev.~Lett.~{\bf 11} (1963) 167. 

 \bibitem{Bar64} V.E. Barnes \EA, Phys.~Rev.~Lett.~{\bf 12} (1964) 204. 

 \bibitem{Gel64b} M. Gell-Mann, Phys.~Lett.~{\bf 8} (1964) 214. 
 
 \bibitem{Zwe64} G. Zweig, CERN preprint 8182/TH401 (1964) unpublished. 
 
 \bibitem{Clo79} F.E. Close, {\it An Introduction to Quarks and Partons}, 
 (Academic Press, New York 1979). 
 
\bibitem{Jon77} L.W. Jones, Rev.~Mod.~Phys.~{\bf 49} (1977) 717. 

\bibitem{Gre64} O.W. Greenberg, Phys.~Rev.~Lett.~{\bf 13} (1964) 598; 
   M.Y Han and Yoichiro Nambu, Phys.Rev.~{\bf 129} (1965) B1006. 

 \bibitem{Hod97} L. Hoddeson, L.M. Brown, M. Riordan and M. Dresden, 
 {\it The Rise of the Standard Model: Particle Physics in the 1960s and 
  1970s}, (Cambridge University Press, 1997). 
 
\bibitem{Aub74} J.J. Aubert \EA, Phys.~Rev.~Lett.~{\bf 33} (1974) 1404; 
  J.E. Augustin \EA, ibid. p. 1406. 

\bibitem{Her77} S.W.Herb \EA, Phys.~Rev.~Lett.~{\bf 39} (1977) 252. 

\bibitem{Abe95} F.Abe \EA, Phys.~Rev.~Lett.~{\bf 74} (1995) 2626. 
  
\bibitem{Isg89} Nathan Isgur and Mark B. Wise, Phys.Lett.~{\bf B232} (1989) 
  113; Phys.Lett.~{\bf B237} (1990) 527. 
  
 \bibitem{Ell96} R.K. Ellis, W.J. Stirling and B.R. Webber, {\it QCD and 
  Collider Physics}, (Cambridge University Press, 1996). 
 
\bibitem{Gro73} D.J.Gross and Frank Wilczek, Phys.~Rev.~Lett.~{\bf 30} (1973) 
  1343; H.D. Politzer, ibid. 1346. 

\bibitem{Ste95} G. Sterman \EA, Rev.Mod.Phys.~{\bf 67} (1995) 157. 

\bibitem{Mat08} N. Mathur, proceedings of this workshop.   
  
\bibitem{F2summ} L.W. Whitlow \EA, (SLAC), Phys.~Lett.~{\bf B282} (1992) 475; 
  C.J. Adloff \EA, (H1), Eur.Phys.J.~{\bf C21} (2001) 33; S. Chekanov \EA, 
  (ZEUS), Eur.Phys.J.~{\bf C21} (2001) 443; A.C. Benvenuti \EA, (BCDMS), 
  Phys.Lett.~{\bf B223} (1989) 485; M. Arneodo \EA, (NMC), 
  Nucl.Phys.~{\bf B483} (1997) 3; M.R. Adams \EA, (E665), Phys.Rev.~{\bf C54} 
  (1996) 3006. 
  
\bibitem{Blo69} E.D. Bloom \EA, Phys.~Rev.~Lett.~{\bf 23} (1969) 930; 
  M. Breidenbach \EA, ibid. p. 935. 
  
\bibitem{G1summ} J. Ashman \EA, (EMC), Nucl.Phys. {\bf B328} (1989) 1; 
  P.L. Anthony \EA, (E142) Phys.Rev. {\bf D54} (1996) 6620; K. Abe \EA, 
 (E143) Phys.Rev. {\bf D58} (1998) 112003; B. Adeva \EA, (SMC) Phys.Rev. 
  {\bf D60} (2000) 072004; A. Airepetian \EA, (HERMES) Phys.Lett. {\bf B442} 
  (1998) 484; K. Abe \EA, (E154) Phys.Rev.Lett. {\bf 70} (1997) 26; 
  P.L. Anthony \EA, (E155) Phys.Lett. {\bf B493} (2000) 19. 

\bibitem{Dok77} Yu.L Dokshitzer, Sov.Phys.JETP~{\bf 46} (1977) 641; 
  V.N. Gribov and L.N.Lipatov, Sov.J.Nucl.Phys.~{\bf 15} (1972) 438;  
  G. Altarelli and G. Parisi, Nucl.Phys.~{\bf B126} (1977) 287. 

\bibitem{Dre70} S.D. Drell and T-M. Yan, Phys.~Rev.~Lett.~{\bf 25} (1970) 316.
  
\bibitem{Mas07} D. Mason, {\it Proceedings of the 14$^{th}$ International 
  Workshop on Deep Inelastic Scattering (DIS06)}, ed. M. Kuze, K. Nagano 
 and K. Tokoshuku (World Scientific Press, Singapore 2007), p. 165.

\bibitem{CTEQ} J.R. Pumplin \EA, (CTEQ), Jour.High 
 Energy~Phys.~{\bf 0202} (2002) 012. 

\bibitem{MRST} A.D. Martin \EA, (MRST), Eur.Phys.J.~{\bf C35} (2004) 325. 

\bibitem{Spe97} J.Speth and A.W. Thomas, Adv.Nuc.Phys. {\bf 24} (1997) 83. 

\bibitem{Lon98} J.T. Londergan and A.W. Thomas, Prog.Part.Nuc.Phys. 
  {\bf 41} (1998) 49. 

\bibitem{Arn94} M. Arneodo, Phys.Rept. {\bf 240} (1994) 301. 

\bibitem{Gee95} D.F. Geesaman, K. Saito and A.W. Thomas, 
  Annu.Rev.Nucl.Part.Sci. {\bf 45} (1995) 337. 

\bibitem{AGK74} V.N. Gribov, arXiv:hep-ph/0006158; V.A. Abramovsky, V.N. 
  Gribov and O.V. Kancheli, Sov.J.Nucl.Phys. {\bf 18} (1974) 308. 

\bibitem{Kul06} S.A. Kulagin and R. Petti, Nucl.Phys.~{\bf A765} (2006) 126.

\bibitem{SMC} B. Adeva \EA, (SMC), Phys.Lett.~{\bf B420} (1998) 180;  
  K. Ackerstaff \EA, (HERMES), Phys.Lett.~{\bf B464} (1999) 123. 

\bibitem{Ans97} E. Leader and M. Anselmino, Z.Phys.~{\bf C41} (1998) 239. 

\bibitem{Bra05} F. Bradamante, Prog.Part.Nucl.Phys.~{\bf 55} (2005) 270. 

\bibitem{Delg} V. Yu Alexakhin \EA (COMPASS), Phys.Lett.~{\bf B647} (2007) 
  8; B. Adeva \EA (SMC), Phys.Rev. {\bf D60} (1999) 072004; A. Airapetian 
  \EA (HERMES), Phys.Rev. {\bf D75} (2007) 012007. 

\bibitem{Age07} E.S. Ageev \EA (COMPASS), Phys.Lett.~{\bf B647} (2007) 330.  

\bibitem{Ji98} X.D. Ji, Phys.Rev.Lett.~{\bf 78} (1997) 610. 

\bibitem{Tho08} A.W. Thomas, proceedings of this workshop.   
  
\bibitem{JLAB} {\it The Science and Experimental Equipment for the 12 GeV 
 Upgrade of CEBAF}, http://www.jlab.org/12GeV/development.html.

\bibitem{Des05} A. Deshpande, R. Milner, R. Venugopalan and W. Vogelsang, 
   Ann.Rev.Nucl.Part.Sci. {\bf 55} (2005) 165. 
  
\bibitem{Mcl94} L.D. McLerran and R. Venugopalan, Phys.~Rev.~{\bf D49}  
  (1994) 2233; ibid. p. 3352. 



\end{thebibliography}

\end{document}